# THE LIGHT VELOCITY CASIMIR EFFECT

Does the Velocity of Light in a Vacuum Increase When Propagating Between the Casimir Plates?

**Tom Ostoma and Mike Trushyk**

48 O'HARA PLACE, Brampton, Ontario, L6Y 3R8
miket1@home.com

Wednesday, November 24, 1999

## ACKNOWLEGMENTS

We wish to thank Paul Almond for originally pointing out this effect to us and for his review and comments on this work. We also want to thank Paul for the many interesting e-mail exchanges on the subject of space, time, light, matter, and CA theory. We thank R. Mongrain for many long discussions on the nature of quantum theory and space-time, and for his insight on the way photons propagate through the quantum vacuum.

# ABSTRACT

*Our theory of quantum gravity called Electro-Magnetic Quantum Gravity (EMQG) depends heavily on an important property of the quantum vacuum; it's ability to effect the velocity of photon propagation under two very special physical conditions. In the first case, photon propagation in the vacuum is altered when the electrically charged virtual particle density changes, as it does between the Casimir plates. In the second case, photon propagation in the vacuum is also altered when there is a coordinated acceleration given to the electrically charged virtual particles of the quantum vacuum, such as near a large mass like the earth (EMQG). These effects can be understood through the familiar process that photons partake when interacting with all electrically charged particles; 'Photon Scattering'.*

*Here we propose experiments that might be set up to detect the increase in the velocity of light in a vacuum in the laboratory frame for the first case, that is when photons travel between (and perpendicular to) the Casimir plates in vacuum. The Casimir plates are two closely spaced, conductive plates, where an attractive force is observed to exist between the plates called the 'Casimir Force'. We propose that the velocity of light in a vacuum increases when propagating between the Casimir Plates, which are in a vacuum. We call this effect the 'Light Velocity Casimir Effect' or LVC effect. In the second case where light propagates upwards or downwards on the earth, the change in light velocity predicted by EMQG is associated with a corresponding curved 4D space-time in general relativity, where light velocity is taken as constant. We find that it is impossible to distinguish between these two conflicting views of light propagation in large gravitational fields by experimental means at this time.*

*The LVC effect happens because the vacuum energy density in between the plates is lower than that outside the Casimir plates. The conductive plates <u>disallow</u> certain frequencies of electrically charged virtual particles to exist inside the plates, thus lowering the inside vacuum particle density, compared to the density outside the plates. The Casimir plates also disallow certain wavelengths of virtual photons as well, which is the basis for the calculation of the Casimir force first done by H.B. G. Casimir in 1948. The reduced (electrically charged) virtual particle density results in fewer photon scattering events inside the plates, which should increase the light velocity between the plates in a vacuum relative to the normal vacuum light speed (as measured with instruments in the laboratory frame). A similar effect, involving light velocity change, happens when light travels through two different real material densities; for example when light propagates from water to air, a process known as optical refraction. We also propose an experiment to demonstrate the Casimir refraction of light moving at a shallow angle that is nearly perpendicular to a series of unequally spaced Casimir plates, which cause a permanent shift in the direction of light propagation. Furthermore we propose a method to determine the index of refraction for light propagating from the ordinary vacuum to the less dense Casimir vacuum.*

# **TABLE OF CONTENTS**

1.     INTRODUCTION

*"The relativistic treatment of gravitation creates serious difficulties. I consider it probable that the principle of the constancy of the velocity of light in its customary version holds only for spaces with constant gravitational potential."*
- Albert Einstein  (in a letter to his friend Laub, August 10, 1911)

The subject of this work might seem like scientific heresy to the reader. No doubt many of you will instantly reject this proposal on the grounds that it violates special relativity. Einstein's 1905 prediction that the speed of light in a vacuum is an absolute constant for all inertial observers has now been well established theoretically and experimentally, although there may still be a few small cracks in the armor as witnessed by quantum non-locality and quantum tunneling effects (section 4). Special relativity has stood the test of time for 95 years without any evidence to the contrary, and has thus become a corner stone of modern theoretical physics. It would seem to be a career ending move for anyone to propose that the light velocity in a vacuum is anything but the accepted fixed value!

Yet we nevertheless propose an experiment which we call the 'Light Velocity Casimir' experiment or the 'LVC' experiment, to prove that the ***velocity of light*** (front velocity) propagating in vacuum inside and perpendicular to two closely spaced, electrically conducting and transparent plates called the Casimir plates, ***will actually increase*** inside the plates compared to the light velocity in the normal vacuum as measured in the laboratory frame (figure 1). The Casimir plates are a pair of closely spaced, conductive plates, where an attractive force is observed to exist between the plates called the 'Casimir Force'. H.B.G. Casimir theoretically predicted the existence of this force in 1948 (ref. 6). Recently the Casimir force has been verified experimentally for a plate spacing of about 1 micrometer, with an accuracy of about 5% (1996, ref. 7), by using an electromagnetic-based torsion pendulum. More recently U. Mohideen and A. Roy have made an even more precise measurement of the Casimir force in the 0.1 to 0.9 micrometer plate spacing to an accuracy of about  1% (1998, ref. 8) using the techniques of atomic force microscopy.

We believe that there must exist a 'Vacuum Casimir Index of Refraction' called '$n_{vac}$' for light traveling from outside, and then through the Casimir plates. The Casimir vacuum index of refraction is defined as the ratio of the velocity of light in normal vacuum conditions ('c' or 299,792.458 km/sec) divided by the light velocity measured perpendicular to the Casimir plates in vacuum ('$c_c$' or slightly greater than 299,792.458 km/sec). The vacuum Casimir index of refraction '$n_{vac}$' is given by: $n_{vac} = c / c_c$ , which is slightly less than one for the Light Velocity Casimir (LVC) experiment. The process is comparable to the familiar index of refraction for light propagating from water to air, where the light velocity in air is greater than the light velocity in water. Figure #2 shows a possible experiment to observe the Casimir vacuum index of refraction by witnessing the deflection of light on a shallow angle propagating through a series of unequally spaced, Casimir plates (described in section 5).

Furthermore we have theoretical reason to believe that the index of refraction $n_{vac}$ will vary with the Casimir plate spacing 'd', just as the Casimir force varies with plate spacing. In the Casimir force effect the force varies as the inverse fourth power of the plate spacing. The light velocity dependence on plate spacing is unknown at this time. Before we go into details on the LVC experiment we explain precisely what we mean by the increase in light velocity in the vacuum.

## 1.1 DEFINITION OF EINSTEIN CAUSALITY

Often in the literature we find statements made regarding Einstein causality, and that nothing propagates faster than the speed of light in the vacuum. Since light consists of photons, which have both particle and wave properties, care must be taken to ensure that velocity of light is properly defined. This is especially important when taking into account the quantum wave packet properties of photons. In L. Brillouin classic book titled 'Wave Propagation and Group Velocity, 1960', he identified five different definitions for the velocity of a finite-bandwidth pulse of electromagnetic radiation. We will be concerned only with the 'front velocity' of light, which is defined below.

The proper definition for light velocity that we use here is the 'front velocity', which was given by Sommerfield and L. Brillouin and (ref. 33). Suppose that there is a light source at point x=0, which is switched on at the time t=0. Some distance 'd' away from the source at x, no effect can be detected that is coming from point x before the time 'd/c'. The beginning of the signal is a discontinuity in the signal envelope (or in a higher derivative). The beginning of the signal is known as the front velocity and this may not exceed the velocity of light in a vacuum in order to fulfill the Einstein causality condition. We maintain that the front velocity for a beam of light traveling through the Casimir plates *will exceed* the velocity of light in the ordinary vacuum during transit through the Casimir plates.

## 1.2 INTRODUCTION TO THE QUANTUM VACUUM LIGHT SCATTERING

How did we come to such a drastic conclusion regarding the increased velocity of light, in spite of the heavy body of scientific evidence to the contrary? This conclusion is partly based on our work on a new quantum theory of gravity called Electro-Magnetic Quantum Gravity (EMQG) ref. 1, which depends heavily on an important characteristic of the quantum vacuum; it's ability to affect the velocity of photon propagation under certain special conditions through the familiar process called 'Photon Scattering'.

EMQG requires a photon scattering process, photon scattering in the accelerated quantum vacuum near the earth, in order to understand gravity and the Newtonian equivalence of inertial mass and gravitational mass on a quantum scale. According to EMQG the light velocity in a region where the virtual particle vacuum density is lower than in the normal vacuum, is greater than the light velocity in the normal vacuum. This occurs for an observer in the laboratory frame using his clocks and rulers to measure the front velocity of light.

According to the general principles of EMQG theory, the LVC effect occurs because:

1. The fundamental virtual matter particles (fermions) that make up the quantum vacuum (not the ZPF or virtual photons as in the Casimir force effect) are ultimately electrically charged at the lowest level, just as we believe for real ordinary matter.
2. Since at the lowest level the virtual particles of the vacuum are electrically charged, they will interact with light (photon particles) propagating through the vacuum.
3. According to quantum theory when a photon interacts with an electrically charged virtual particle, the propagation is delayed at each electrically charged virtual particle of the quantum vacuum, before the photon continues propagating. *Why is there a photon delay?* There is a time delay during the absorption and subsequent re-emission of the photon by a given charged virtual particle. The uncertainty principle places a lower limit on this time delay, and forbids it from being zero. In other words, according to the uncertainty principle the time delay due to the absorption and re-emission time of the photon cannot be exactly equal to zero.
4. The time delay caused by the absorption and subsequent re-emission of the photon by a given electrically charged virtual particle results in a **lower TOTAL** AVERAGE velocity for the propagation of the photons on the macroscopic distance scale, as compared to the average velocity of the photon without the presence of any constraining Casimir plates.
5. In the case of the vacuum between the Casimir plates, the virtual *fermion particle density* in the vacuum is lower between the plates compared to that outside the plates. The Casimir plates prohibit certain wavelengths of the fermion wave function from existing (as it does for certain photon wavelengths in the Casimir force calculation).
6. Therefore, the **TOTAL AVERAGE light velocity** inside the plates must be **greater** than that outside the plates.

Our review of the physics literature has not revealed any previous work on the time delay analysis of photon propagation through the ordinary quantum vacuum or any evidence to contradict our hypothesis of photon vacuum delay, presumably because of the precedent set by Einstein's postulate of light speed constancy.

**Note:** *Suppose we can place a tiny observer A and his clocks and rulers (somehow) in between the Casimir plates. We would find for observer A that **his measurement** of the speed of light is the **same** as the conventional value! However for an observer B outside the plates in the laboratory frame, his measurement **does** show an increase in light velocity through the plates. Furthermore, the individual space and time measurements '$d_{lab}$' and '$t_{lab}$' made by observer B in the laboratory frame **do not agree** with the same measurements of '$d_{plate}$' and '$t_{plate}$' made by our tiny observer A inside the Casimir plates. The general relativists could argue that the 4D space-time inside the Casimir plates is altered compared to outside, and that the light velocity is still an absolute constant in all cases! This argument results because of the crucial importance of light propagation to*

*the fundamental nature of space and time measurements, a theme that was first championed by Einstein. This same controversy rears it's ugly head in gravity where we are forced to choose between the two experimentally indistinguishable views; 4D curved space-time in general relativity and variable light velocity proposal of EMQG. It turns out to be impossible to distinguish between curved 4D space-time in gravitational frames, and variations in light velocity in gravitational frames experimentally. More will be said on this important point in appendix A (section A10 and ref. 1).*

Another result from EMQG states that the velocity of light without the existence of *any* of the virtual particles of the quantum vacuum ought to be *much greater* than the observed average light velocity in the vacuum. The electrically charged virtual fermion particles of the quantum vacuum frequently scatter photons, which introduce many tiny delays for the photon propagation. This causes a great reduction in the total average light velocity in the vacuum that is populated by countless numbers of virtual, electrically charged, particles. In other words; ***the low level light velocity (between virtual particle scattering events) is much greater than the measured average light velocity after vacuum scattering in the normal vacuum***. A similar effect is known to occur when light propagates through glass, where photons scatter with electrons in the glass molecules, which subsequently reduces the average light velocity through glass compared to the normal vacuum light velocity.

Here we propose an experiment that might be able to detect this predicted increase in the velocity of light between the Casimir plates in vacuum. The experiment is designed to compare the velocity of light in the ordinary vacuum against the light velocity that is propagating in between the Casimir plates, also in a vacuum state (figure 1). A light source is split into two parallel paths by a beam splitter; one path is the reference vacuum path of light, and in the other path the light is allowed to propagate in vacuum perpendicular to two closely spaced, electrically conductive and transparent plates. The two paths are then recombined by a beam splitter, and routed to an interferometer. After the apparatus is calibrated (and if our prediction is correct), the phase of the interference pattern will indicate that light travels faster in the Casimir plate leg of the interferometer. Furthermore, it might be possible to measure the index of refraction for light traveling from the normal vacuum to the Casimir vacuum.

As we have said the light velocity increases between the Casimir plates because the vacuum energy-density in between the plates is lower than that outside the plates. This difference in vacuum energy-density or 'pressure' is actually the cause of the attractive force between the plates, where the energy-density is greater outside. This implies that it might be possible to extract a virtually unlimited supply of energy the quantum vacuum, which is an active area of recent research (ref. 11). Before we go into detail on the proposed LVC experiments, we present a brief introduction to the virtual particles of the quantum vacuum that is crucial to the understanding of the LVC effect.

## 2. THE VIRTUAL PARTICLES OF THE QUANTUM VACUUM

*Philosophers:    "Nature abhors a vacuum."*

In order to make a complete vacuum, one must remove all matter from an enclosure. However one would find that this is still not good enough. One must also lower the temperature inside the closure to absolute zero in order to remove all thermal electromagnetic radiation. Nernst correctly deduced in 1916 (ref. 32) that empty space is still not completely devoid of all radiation after this is done. He predicted that the vacuum is still permanently filled with an electromagnetic field propagating at the speed of light, called the zero-point fluctuations (or sometimes called by the generic name 'vacuum fluctuations'). This result was later confirmed theoretically by the newly developed quantum field theory that was developed in the 1920's and 30's.

Later with the development of QED (the quantum theory of electrons and photons), it was realized that all quantum fields should contribute to the vacuum state. This means that virtual electrons and positron particles should not be excluded from consideration. These particles possess mass and have multiples of half integer spin (such as the electron), and therefore belong to the generic class of particles known as fermions. We refer to virtual electrons and virtual anti-electrons (positron) particles as virtual fermions. We believe that ultimately all fermions can be broken down to a fundamental entity that is also electrically charged, as well as having half integer spin and mass (technically, mass-charge as described in Appendix A).

According to modern quantum field theory, the perfect vacuum is teeming with activity as all types of quantum virtual particles (and virtual bosons or the force carrying particles) from the various quantum fields appear and disappear spontaneously. These particles are called 'virtual' particles because they result from quantum processes that generally have short lifetimes, and are mostly undetectable. One way to look at the existence of the quantum vacuum is to consider that quantum theory forbids the complete absence of propagating fields. This is in accordance with the famous Heisenberg uncertainty principle. In general, it is known that all the possible real particles types (for example electrons, quarks, etc.) will also be present in the quantum vacuum in their virtual particle form.

In the QED vacuum, the quantum fermion vacuum is produced from the virtual particle pair creation and annihilation processes that create enormous numbers of virtual electron and virtual positron pairs. We also have in QED the creation of the zero-point-fluctuation (ZPF) of the vacuum consisting of the electromagnetic field or virtual photon particles. Indeed in the standard model, we also find in the vacuum every possible boson particle, such as the gluons, gravitons, etc., and also every possible fermion particle, such as virtual quarks, virtual neutrinos, etc.

2.1  INTRODUCTION TO THE CASIMIR FORCE EFFECT

The existence of virtual particles of the quantum vacuum reveals itself in the famous Casimir effect (ref. 6), which is an effect predicted theoretically by the Dutch scientist Hendrik Casimir in 1948. The Casimir effect refers to the tiny attractive force that occurs between two neutral metal plates suspended in a vacuum. He predicted theoretically that the force 'F' per unit area 'A' for plate separation D is given by:

$$F/A = -\pi^2 h c /(240 D^4) \quad \text{Newton's per square meter} \quad \text{(Casimir Force 'F')} \quad (2.1)$$

Casimir obtained this formula by calculating the sum of the quantum-mechanical zero-point energies of the normal modes of the electromagnetic field (virtual photons) between two conductive plates.

The origin of this minute force can be traced to the disruption of the normal quantum vacuum virtual photon distribution between two nearby metallic plates as compared to the vacuum state outside the plates. Certain virtual photon wavelengths (and therefore energies) are forbidden to exist between the plates, because these waves do not 'fit' between the two plates (which are both at a relative classical electrical potential of zero). This creates a negative pressure due to the unequal energy distribution of virtual particles inside the plates as compared to those outside the plate region. The pressure imbalance can be visualized as causing the two plates to be drawn together by radiation pressure. Note: Even in the vacuum state, the virtual photon particles *do* carry energy and momentum while they exist.

Although the Casimir effect has been attributed to the zero-point fluctuations (ZPF) in the EM field inside the plates, Schwinger showed in the late 70's that the Casimir effect can also be derived in terms of his source theory (ref. 13), which has no explicit reference to the ZPF of the EM field between the plates. Recently Milonni and Shih have developed a theory of the Casimir force effect, which is totally within the framework of conventional QED (ref. 15). Therefore it seems that it is only a matter of taste whether we attribute the Casimir force effect to the ZPF fields or to the matter fields in vacuum (ref. 23).

Recently Lamoreaux made accurate experimental measurements for the first time of the Casimir force existing between two gold-coated quartz surfaces that were spaced on the order of a micrometer apart (ref. 7). Lamoreaux found a force value of about 1 billionth of a Newton, agreeing with the Casimir theory to within an accuracy of about 5%. More recently, U. Mohideen and A. Roy have made an even more precise measurement in the 0.1 to 0.9 micrometer plate spacing to an accuracy of about 1% (1998, ref. 8). Therefore the experimental reality of this effect is beyond question.

Can the vacuum state be disrupted by other physical processes besides the Casimir plates? One might ask what happens to the virtual particles of the quantum vacuum that are subjected to a large gravitational field like the earth? Since the quantum vacuum is composed of virtual fermions (as well as virtual bosons), the conclusion is inescapable: ***all*** *the virtual fermions possessing mass must be falling (accelerating) on the average towards the earth during their very brief lifetimes*. This vacuum state is definitely different from the vacuum of far outer space away from gravitational fields. Yet to our knowledge, no previous authors have acknowledged the existence of this effect, or studied the physical consequences that result from this. It turns out that the free fall state of the virtual, electrically charged fermion particles of the vacuum is actually the root cause of 4D space-time curvature and also leads to a full understanding of the principle of equivalence. In EMQG (appendix A) we fully study the consequences of a falling quantum vacuum in quantum gravity, which does lead to new experimentally testable predictions.

The physics of the Casimir force effect implies that the quantum vacuum contains an enormous reservoir of energy (ref. 11). Although in standard quantum field theory the number density of virtual particles is unlimited, some estimates place a high frequency cut-off at the plank scale which is estimated to be a density of $10^{90}$ particles per cubic meter (ref. 11)! Generally this energy-density is not available because the energy-density is uniform and it permeates everything. It's like the situation in the deep ocean, where deep sea fishes easily tolerate the extreme pressures in the abyss, because the pressure inside and outside the fish's body balance. If a human goes into these depths, a great difference in pressure must be maintained to support atmospheric pressure inside the human body. Some physicists are looking at ways in which this vast energy reservoir can be tapped (ref. 11)

If the vacuum is capable of exerting a mechanical force between the two Casimir plates, might the vacuum's effect be felt in a less exotic way? Most physicists believe that the answer is no. Yet there is a small number of physicist who believe otherwise. In 1994, R. Haisch, A. Rueda, and H. Puthoff (ref. 5) were the first to propose a theory of inertia (known here as HRP Inertia), where the quantum vacuum played a central role in Newtonian inertia. They suggested that inertia is due to the strictly local electrical force interactions of charged matter particles with the immediate background virtual particles of the quantum vacuum. We have built on their work and developed a theory of quantum gravity and quantum inertia based on their idea. According to EMQG, the quantum vacuum affects all masses that are in the state of acceleration. In the EMQG model, the force of inertia is actually caused by the resistance force to acceleration by the electrical force interactions between charged particles that make up a mass and the electrically charged virtual particles of the quantum vacuum. We call this quantum inertia, which plays a central role in our quantum theory of gravity that closely links inertia and gravity. We introduce this important concept in section 2.3 (this section can be omitted if desired, since it is not essential in order to understand the LVC experiments).

## 2.2 EVIDENCE FOR THE EXISTENCE OF VIRTUAL PARTICLES (** Optional)

There is other evidence for the existence of virtual particles besides the Casimir force effect. We present a very brief review of some theoretical and experimental evidence for the existence of the virtual particles of the quantum vacuum:

(1) The extreme precision in the theoretical calculations of the hyper-fine structure of the energy levels of the hydrogen atom, and the anomalous magnetic moment of the electron and muon are both based on the existence of virtual particles in the framework of QED. These effects have been calculated in QED to a very high precision (approximately 10 decimal places), and these values have also been verified experimentally to an unprecedented accuracy. This indeed is a great achievement for QED, which is essentially a perturbation theory of the electromagnetic quantum vacuum. Indeed, this is one of physics greatest achievements.

(2) Recently, vacuum polarization (the polarization of electron-positron pairs near a real electron particle) has been observed experimentally by a team of physicists led by David Koltick. Vacuum polarization causes a cloud of virtual particles to form around the electron in such a way as to produce an electron charge screening effect. This is because virtual positrons tend to migrate towards the real electron, and the virtual electrons tend to migrate away. A team of physicists fired high-energy particles at electrons, and found that the effect of this cloud of virtual particles was reduced the closer a particle penetrated towards the electron. They reported that the effect of the higher charge for the penetration of the electron cloud with energetic 58 giga-electron volt particles was equivalent to a fine structure constant of 1/129.6. This agreed well with their theoretical prediction of 128.5 of QED. This can be taken as verification of the vacuum polarization effect predicted by QED, and further evidence for the existence of the quantum vacuum.

(3) The quantum vacuum explains why cooling alone will never freeze liquid helium. Unless pressure is applied, vacuum energy fluctuations prevent its atoms from getting close enough to trigger solidification.

(4) For fluorescent strip lamps, the random energy fluctuations of the virtual particles of the quantum vacuum cause the atoms of mercury, which are in their exited state, to spontaneously emit photons by eventually knocking them out of their unstable energy orbital. In this way, spontaneous emission in an atom can be viewed as being directly caused by the state of the surrounding quantum vacuum.

(5) In electronics, there is a limit as to how much a radio signal can be amplified. Random noise signals are always added to the original signal. This is due to the presence of the virtual particles of the quantum vacuum as the real radio photons from the transmitter propagate in space. The vacuum fluctuations add a random noise pattern to the signal by slightly modifying the energy of the propagating radio photons.

(6) Recent theoretical and experimental work done in the field of Cavity Quantum Electrodynamics suggests that the orbital electron transition time for excited atoms can be affected by the state of the virtual particles of the quantum vacuum immediately surrounding the excited atom in a cavity, where the size of the cavity modifies the spectrum of the virtual particles.

In the weight of all this evidence, only a few physicists doubt the existence of the virtual particles of the quantum vacuum. Yet to us, it seems strange that the quantum vacuum should barely reveal it's presence to us, and that we only know about it's existence through rather obscure physical effects like the Casimir force effect and Davies-Unruh effect. This is especially odd considering that the observable particles of ordinary real matter in an average cubic meter of space in the universe constitute a minute fraction of the total population of virtual particles of the quantum vacuum at any given instant of time.

Some estimates of the quantum vacuum particle density (ref. 11) place the vacuum particle numbers at about $10^{90}$ particles per cubic meter! Instead, we believe that the quantum vacuum plays a much more prominent role in physics. We maintain that the effects of the quantum vacuum are present in virtually all physical activity. In fact, Newton's three laws of motion can be understood to originate directly from the effects of the quantum vacuum (Appendix A). Furthermore, the quantum vacuum plays an extremely important role in gravity, which is generally well understood by the physics community.

In order not to distract the reader from the main theme of this paper, we have included a brief review of EMQG theory which summarizes the central role that light scattering in the accelerated quantum vacuum has in our quantum gravity theory; and for the principle of equivalence, inertia, and 4D space-time curvature. This can be found in Appendix A of this paper. A full account is given in reference 1. We provide an optional introduction to Quantum Inertia in the next section for those readers who are interested.

2.3     INTRODUCTION TO QUANTUM INERTIA THEORY (** Optional)

Recently it has been proposed that Newtonian Inertia is strictly a quantum vacuum phenomenon! If this is true, then the existence of the quantum vacuum actually reveals it's presence to us in all daily activities! Unlike the hard-to-measure Casimir force effect, the presence of the inertial force is universal and it's presence prevails throughout all of physics. For example, the orbital motion of the earth around the sun is a balancing act between inertia force and gravitational force. If quantum inertia is true, every time you accelerate, you are witnessing a quantum vacuum effect! This is a far cry from an exotic and almost impossible measurement of the feeble Casimir force between two plates.

In 1994, R. Haisch, A. Rueda, and H. Puthoff (ref. 5) were the first to propose a theory of inertia (known here as HRP Inertia), where the quantum vacuum played a central role in Newtonian inertia. They suggested that inertia is due to the strictly local electrical force interactions of charged matter particles with the immediate background virtual particles of the quantum vacuum (in particular the virtual photons or ZPF as the authors called it). They found that inertia is caused by the magnetic component of the Lorentz force, which arises between what the author's call the charged 'parton' particles in an accelerated reference frame interacting with the background quantum vacuum virtual particles. The sum of all these tiny forces in this process is the source of the resistance force opposing accelerated motion in Newton's F=MA. The 'parton' is a term that Richard Feynman coined for the constituents of the nuclear particles such as the proton and neutron (now called quarks).

We have found it necessary to make a small modification to HRP Inertia theory as a result of our investigation of the principle of equivalence. Our modified version of HRP inertia is called "Quantum Inertia" (or QI), and is described in detail in Appendix A. This theory also resolves the long outstanding problems and paradoxes of accelerated motion introduced by Mach's principle, by suggesting that the vacuum particles themselves serve as Mach's universal reference frame (for <u>acceleration</u> only), without violating the principle of relativity of constant velocity motion. In other words our universe offers no observable reference frame to gauge inertial frames (non-accelerated frames where Newton's laws of inertia are valid), because the quantum vacuum offers no means to determine absolute constant velocity motion. However for accelerated motion, the quantum vacuum plays a very important role by offering a resistance to acceleration, which results in an inertial force opposing the acceleration of the mass. Thus the very existence of inertial force reveals the absolute value of the acceleration with respect to the net statistical average acceleration of the virtual particles of the quantum vacuum. If this is correct then Newton's three famous laws of motion can be understood at the quantum level (ref. 20).

There have been various clues to the importance the virtual particles of the quantum vacuum for the accelerated motion of real charged particles. One example is the so-called Davies-Unruh effect (ref. 18), where uniform and linearly accelerated charged particles in the vacuum are immersed in a heat bath, with a temperature proportional to acceleration (with the scale of the quantum heat effects being very low). However, the work of reference 5 is the first place we have clearly seen the identification of inertial forces as the direct consequence of the interactions of real matter particles with the quantum vacuum.

It has also even been suggested that the virtual particles of the quantum vacuum are somehow involved in gravitational interactions. The prominent Russian physicist A. Sakharov proposed in 1968 (ref. 16) that Newtonian gravity could be interpreted as a van der Waals type of force induced by the electromagnetic fluctuations of the virtual particles of the quantum vacuum. Sakharov visualized ordinary neutral matter as a collection of electromagnetically, interacting polarizable particles made of charged point-mass sub-particles (partons). He associated the Newtonian gravitational field with the Van Der Waals force present in neutral matter, where the long-range radiation fields are generated by the parton 'Zitterbewegung'. Sakharov went on to develop what he called the 'metric elasticity' concept, where space-time is somehow identified with the 'hydrodynamic elasticity' of the vacuum. However, he did not understand the important clues about the quantum vacuum that are revealed by the equivalence principle, nor the role that the quantum vacuum played in inertia and Mach's principle. We maintain that the quantum vacuum also make it's presence felt in a very *big* way in all gravitational interactions (Appendix A) just as it does in inertia!

There have been further hints that the quantum vacuum is involved in gravitational physics. In 1974 Hawkings (ref. 17) announced that black holes are not completely black. Black holes emit an outgoing thermal flux of radiation due to gravitational interactions of the black hole with the virtual particle pairs created in the quantum vacuum near the event horizon. At first sight the emission of thermal radiation from a black hole seems paradoxical, since nothing can escape from the event horizon. However the spontaneous creation of virtual particle and anti-particle pairs in the quantum vacuum near the event horizon can be used to explain this effect (ref. 18). Heuristically one can imagine that the virtual particle pairs (that are created with wavelength λ that is approximately equal to the size of the black hole) 'tunnel' out of the event horizon. For a virtual particle with a wavelength comparable to the size of the hole, strong tidal forces operate to prevent re-annihilation. One virtual particle escapes to infinity with positive energy to contribute to the Hawking radiation, while the corresponding antiparticle enters the black hole to be trapped forever by the deep gravitational potential. Thus the quantum vacuum is important in order to properly understand the Hawking radiation.

As a result of all these and other considerations, we have developed a new approach to the unification of quantum theory with general relativity referred to as Electro-Magnetic Quantum Gravity or EMQG (ref. 1 and summary in appendix A). EMQG had its early origins in Cellular Automata (CA) theory (ref. 2,4,9 and 34), and on a theory of inertia proposed by R. Haisch, A. Rueda, and H. Puthoff (ref. 5). In EMQG, the quantum vacuum plays an *extremely* important, if not a central role, in *both* inertia and gravitation. It also plays a *major* role in the origin of 4D curved space-time curvature near gravitational sources.

We maintain that anybody who believes in the existence of the virtual particles of the quantum vacuum and accepts the fact that many virtual particles carry mass (virtual fermions), will have no trouble in believing that the virtual particles of the vacuum are falling in the presence of a large gravitational mass like the earth during their brief lifetimes. We believe the existence of the downward accelerating virtual particles, under the action of a large gravitational field, turns out to be the *missing link* between inertia and gravity. It leads us directly to a full understanding of the principle of equivalence. Although the quantum vacuum has been studied in much detail in the past, to our knowledge no one has examined the direct consequences of a quantum vacuum in a state of free-fall near the earth. This concept is the central theme behind EMQG. Reference 14 offers an excellent introduction to the motion of matter in the presence of the quantum vacuum, and on the history of the discovery of the virtual particles of the quantum vacuum.

We propose that the virtual particles of the quantum vacuum can be viewed as kind of a transparent fluid medium, sort of like a kind of a 21$^{th}$ century "ether". Unlike ordinary transparent fluids like water, the vacuum does not resist constant velocity motion. However the virtual particles of the quantum vacuum can be made to take on a coordinated (average) accelerated motion with respect to an observer in two different physical instances, and this has very important consequences for mass particles in the following two cases:

(1) The quantum vacuum looks disturbed from the perspective of a mass that is being accelerated (by a rocket for example). Here the observer and his mass are the physical entities that are actually accelerating, and the quantum vacuum only *appears* to be accelerated in the reference frame of the observer. Here the vacuum acceleration is ***not actually real***! However, the quantum vacuum effects are very real, and are the root cause of inertial force.
(2) The vacuum actually is disturbed by the presence of a near-by gravitational field of a large mass, like the earth. In this case, the coordinated vacuum particle (net average) acceleration with respect to the earth's surface ***is real***, and is caused by direct graviton exchanges between the earth and the individual virtual fermion particles of the quantum vacuum. Therefore, the vacuum fluid can be viewed as falling just as the Niagara Water Falls. However the vacuum fluid does not accumulate at the earth's surface as a real liquid water fluid might, because the virtual particles are short lived and are constantly being replaced by new ones.

These considerations imply that Newton's principle of equivalence of gravitational mass and inertial mass can be understood to be caused by the virtual particles of the quantum vacuum. The inertial mass 'm' is defined in the formula $F = ma$, and has the same magnitude as the gravitational mass 'm' defined in $F = GmM_e/r^2$, which are two *independent* definitions for the same mass value. Newton equivalence results because:

***The quantum vacuum <u>looks the same</u> from the perspective of an accelerated mass 'm' on the floor of a rocket accelerated at 1g, as it does from the perspective of a stationary mass 'm' on the earth!***

In order to see how the Newtonian Equivalence principle of inertia and gravitational mass follows from the accelerated quantum vacuum effects, we only have to recall our Quantum Inertia principle:

***The cause of inertia is the electrical resistance force that appears between the electrically charged, real matter particles that constitute a mass, and the surrounding electrically charged, virtual fermions of the quantum vacuum, where there exists a <u>state of relative acceleration</u> between the real and virtual particle species.***

In other words, an accelerated mass feels the inertial force from the sum of the tiny electrical forces that originate from each electrically charged particle that make up a mass.

Similarly the gravitational mass of the same object, stationary on the earth's surface, also feels the exact same sum of the tiny electrical forces that originate from each electrically charged particle that make up a mass, where now it's the ***virtual particles*** of the quantum vacuum that ***accelerates downwards.*** What is the cause of the vacuum particle acceleration on the earth? According to EMQG, which is a quantum field theory of gravity, it is the graviton exchanges between the fermion particles of the earth and the virtual fermion particles of the quantum vacuum. These ideas are fully elaborated in EMQG theory in Appendix A (attached). We now review the basic notions of photon scattering in real matter and in the quantum vacuum.

## 3. LIGHT SCATTERING THEORY

Since photon scattering is essential to our model of the predicted LVC effect (and in EMQG theory), we will examine the general principles of photon scattering in some detail. First we review the conventional physics of light scattering in a ***real media*** such as water or glass, including the concept of the index of refraction and Snell's Law of refraction. We also introduce photon scattering when the real media is moving at a constant velocity, where the velocity of light varies in the moving media and known as the Fizeau effect. Next we introduce an accelerated medium for the real medium and examine how the photons scatter. This is important to understanding EMQG theory. Readers that are only interested in the LVC affect can skip sections 3.3, 3.5 and 3.6. We generalize these arguments to examine photon scattering with the electrically charged virtual particles of the quantum vacuum.

### 3.1 CLASSICAL SCATTERING OF PHOTONS IN REAL MATTER

It is a well-known result of classical optics that light moves slower in glass than in air. Furthermore it is recognized that the velocity of light in air is slower than that of light's vacuum velocity. This effect is described by the index of refraction 'n', which is the ratio of light velocities in the two different media. The Feynman Lectures on Physics gives one of the best accounts of the classical theory for the origin of the refractive index and the slowing of light through a transparent material like glass (ref. 42, chap. 31 contains the mathematical details).

When light passes from a vacuum into glass, with an incident angle of $\theta_0$ it deflects and changes it's direction and moves at a new angle $\theta_1$, where the angles follow Snell's law:

$n = \sin \theta_0 / \sin \theta_1$  (3.11)

This follows geometrically because the wave crests on both sides of the surface of the glass must have the same spacing, since they must travel together (ref. 42). The shortest distance between crests of the wave is the wavelength divided by the frequency. On the vacuum side of the glass surface it is $\lambda_0 = 2\pi c/\omega$, and on the other side it is given by $\lambda =$ $2\pi v/\omega$ or $2\pi c / \omega n$ since we define $v=c/n$. If we accept this, then Snell's law follows geometrically (ref. 42). In some sense, the existence of the index of refraction in Snell's law is confirmation of the change in light speed going from the vacuum to glass.

Snell's law does not tell us why we have a change in light velocity, nor does it give us any insight into the phenomena of dispersion and back scattering of light in refraction. A good classical account of the derivation of the index of refraction is given by Feynman himself in ref. 42. Feynman derives the index of refraction for a transparent medium by accepting that the total electric field in any physical circumstance can be represented by the sum of the fields from all charge sources, and by accepting that the field from a single charge is given by it's acceleration evaluated with a retardation speed 'c' (the propagation speed of the exchanged photons). We only summarize the important points of his argument below, and the full details are available in reference 42:

(1) The incoming source electromagnetic wave (light) consists of an oscillating electric and magnetic field. The glass consists of electrons bound elastically to the atoms, such that if a force is applied to an electron the displacement from its normal position will be proportional to the force.
(2) The oscillating electric field of the light causes the electron to be driven in an oscillating motion, thus acting like a new radiator generating a new electromagnetic wave. This new wave is always delayed, or retarded in phase. These delays result from the time delay required for the bound electron to oscillate to full amplitude. Recall that the electron carries mass and therefore inertia. Therefore some time is required to move the electron.
(3) The total resulting electromagnetic wave is the sum of the source electromagnetic wave plus the new phase-delayed electromagnetic wave, where the total resulting wave is phase-shifted.
(4) The resulting phase delay of the electromagnetic wave is the root cause of the reduced velocity of light observed in the medium.

Feynman goes on to derive the classic formula for the index of refraction for atoms with several different resonant frequency $\omega_k$ which is given by:

$$n = 1 + [q_e^2 / (2e_0 m)] \sum_k N_k / [\omega_k^2 - \omega^2 + i\gamma_k \omega] \qquad (3.12)$$

where n is the index of refraction, $q_e$ is the electron charge, m is the electron mass, $\omega$ is the incoming light frequency, $\gamma_k$ is the damping factor, and $N_k$ is the number of atoms per unit volume. This formula describes the index of refraction for many substances, and also describes the dispersion of light through the medium. Dispersion is the phenomenon where the index of refraction of a media varies with the frequency of the incoming light, and is the reason that a glass prism bends light more in the blue end than the red end of the spectrum.

If the medium consists of free, unbound electrons in the form of a gas such as in a plasma (or as the conduction electrons in a simple metal) then the index of refraction with the conditions $\gamma_k << \omega$ and $\omega_k = 0$ is given by (ref. 42):

$$n \approx 1 - [N_k q_e^2 / (2e_0 m)] / \omega^2 \approx \{1 - [N_k q_e^2 / (e_0 m)] / \omega^2\}^{1/2} \qquad (3.13)$$

where we recall that $(1-x)^{1/2} \approx 1 - x/2$ if x is much lees than 1.

The quantity $\Omega = [N_k q_e^2 / (e_0 m)]^{1/2}$ is sometimes called the Plasma frequency $\Omega$, where there is a transition to the transparent state at $\Omega = \omega$.

## 3.2    QUANTUM FIELD THEORY OF PHOTON SCATTERING IN MATTER

Although the classical account of scattering predicts the experimentally confirmed results, the correct account must be a quantum mechanical account. To quote R. Feynman himself:                   " … yes, but the world is quantum not classical *dam-it*".

The propagation of light through a transparent medium is a very difficult subject in QED. It is impossible to compute the interaction of a collection of atoms with light exactly. In fact, it is impossible to treat even one atom's interaction with light exactly in QED. However the interaction of a real atom with photons can be approximated by a simpler quantum system. Since in many cases only two atomic energy levels play a significant role in the interaction of the electromagnetic field with atoms, the atom can be represented by a quantum system with only two energy eigenstates.

In the text book "Optical Coherence and Quantum Optics" a thorough treatment of the absorption and emission of photons in two-level atoms is given (ref. 43, Chap. 15, pg. 762). When a photon is absorbed, and later a new photon of the same frequency is re-emitted by an electron bound to an atom, there exists a time delay before the photon re-emission. The probabilities for emission and absorption of a photon is given as a function of time $\Delta t$ for an atom frequency of $\omega_0$ and photon frequency of $\omega_l$ :

Probability of Photon Absorption is:   $K [\sin(0.5(\omega_l - \omega_0) \Delta t) / (0.5(\omega_l - \omega_0))]^2$
Probability of Photon Emission is:     $M [\sin(0.5(\omega_l - \omega_0) \Delta t) / (0.5(\omega_l - \omega_0))]^2$
(3.21)
(where K and M are complex expressions defined in ref. 43)

The important point we want to make from eq. 3.21 is that the probability of absorption or emission depends on the length of time $\Delta t$, where the probability of the emission is zero, if the time $\Delta t = 0$. In other words according to QED, a ***finite time*** is required before re-emission of the photon. There are other factors that affect the probability, of course. For example, the closer the frequency of the photon matches the atomic frequency, the higher the probability of re-emission in some given time period. We maintain that these delays are the actual route cause of the index of refraction in a medium.

We believe that a similar thing happens when photons propagate through the quantum vacuum. Therefore, we want to address the effect of the virtual particles of the quantum vacuum on the propagation velocity of real (non-virtual) photons, a subject that is largely ignored in the physics literature.

## 3.3    THE SCATTERING OF PHOTONS IN THE QUANTUM VACUUM

In section 3.1 we discussed photon scattering in a real matter medium and in a real negatively charged electron gas. The electron gas model is the closest model we have towards understanding the Casimir index of refraction of the quantum vacuum. However, there are several important differences between the charged electron gas medium and the electrically charged virtual fermion particles of the quantum vacuum as a medium.

First, and most importantly, virtual particles do not carry any net average energy. Instead an individual virtual particle 'borrows' a small amount of energy during it's brief existence, which is then paid back quickly in accordance to the uncertainty principle. It is because quantum mechanics forbids knowing the value of two complementary variables precisely (in this case energy and existence time) for a virtual particle that virtual particles are allowed to exist at all. Therefore unlike the electron gas, the vacuum is incapable of permanently absorbing light that propagates through it.

Thus the quantum vacuum does not absorb any light over macroscopic distance scales. This statement seems trivial, but it is never-the-less important when considering the quantum vacuum as a medium. On microscopic scales real photons are absorbed and re-emitted by individual virtual particles, in accordance with QED. Photon energy is lost in some collisions and regained in others so that on the average the energy loss is zero. This is because during the brief existence time of a virtual fermion particle, the virtual particle *does* possess energy, which is paid back almost immediately. This quantum process happens an enormous number of times as light travels through macroscopic distance scales. The energy balances out to zero over sufficiently large distance scales.

Furthermore unlike the electron gas, there can be no dispersion of light in the quantum vacuum. In other words all frequencies of electromagnetic radiation move at the same speed through the quantum vacuum in spite of the incredible numbers of virtual particle interactions that occur for any particular frequency of photon. Zero dispersion follows experimentally from many astronomical observations of distant supernova, where there is a dramatic change in light and electromagnetic radiation with time. Observations have been made of specific events in the light curves of supernovae light curves that range from the radio band frequencies to the X-ray / Gamma Ray frequencies. All the different frequencies are observed to arrive on the earth at the same time.

With distances of thousands or millions of light years away, any discrepancy in the photon velocity of supernovae at different frequencies would be very apparent. For example with the relatively nearby supernova 1987A (which exploded about 160,000 years ago in the Large Magellanic cloud) all the different frequencies of EM waves has reached us very much at the same time. If there had been a dispersion of only 0.01 m/sec in light velocity (i.e. 3 parts in $10^{-11}$) between two different frequencies, then the light of one frequency would arrive on the earth:

160000 x 365.25 x 24 x 60 x 60 x $10^{-2}$ / ($3 \times 10^{-8}$) =170 seconds or 2.8 minutes later!

A result like this obviously disagrees with observations made of the spectrum of supernova 1987A. Spectra have been obtained for very distant supernovae up to a few billion light years away in other galaxies. One study places the maximum allowed dispersion to be on the order of 1 part in $10^{-21}$ (ref. xx). Thus we conclude that there is no dispersion of light in the vacuum.

Is there a possibility for an index of refraction in the vacuum, as we have in an electron gas? Remember that an index of refraction requires two different media in which to compare the relative velocities of light. However the vacuum particle density must nearly uniform, with no transitions in density. Let us imagine a situation where somehow we have removed all the virtual particles in half of an empty box in vacuum, and the other half has the normal population of virtual particles in the normal quantum vacuum state. Would there be an index of refraction as light traveled from one side of the box to the other?

This is a very important question because the validity of special relativity at the sub-microscopic distance scales comes into question here. You might think that if the vacuum has no energy, there should no effect on the propagation speed of photons. However we believe that the virtual particles in the quantum vacuum *do* indeed delay the progress of photons through electrically charged vacuum particle scattering effects. Thus we believe that photon scattering reduces the light velocity on the half of the box with electrically charged virtual particles. How can we justify this belief, in spite of the contradiction to special relativity? Special relativity is a classical theory, and was developed in the macroscopic domain of physics. It is almost impossible to measure light velocities over the extremely short distance scales that we are talking about.

The electrically charged virtual particles in the quantum vacuum all have random velocities and move in random directions. They also have random energies $\Delta E$ during their brief life time $\Delta t$, which satisfies the uncertainty principle: $\Delta E \, \Delta t > h/(2\pi)$. Imagine a real photon propagating in a straight path through the electrically charged virtual particles in a given direction. The real photon will encounter an equal number of virtual particles moving towards it as it does moving away from it. The end result is that the electrically charged quantum vacuum particles do not contribute anything different than the situation where *all* the virtual particles in the it's path were at relative rest. Thus we can consider the vacuum as some sort of stationary crystal medium of virtual particles with a very high density, where each virtual particle is short-lived and constantly replaced (and carry no net average energy as discussed above).

The progress of the real photon is delayed as it travels through this quantum vacuum 'crystal', where it meets uncountable numbers of electrically charged virtual particles. Light travels through this with no absorption or dispersion. Based on our general arguments in the sections 3.1 to 3.4 above, we *postulate* that the photon is delayed as it travels through the quantum vacuum. We can definitely say that the uncertainty principle places a lower limit on the emission and absorption time delay, and forbids the time delay from being *exactly* equal to zero.

***Therefore we conclude that the electrically charged virtual particles of the quantum vacuum frequently absorb and re-emit the real photons moving through the vacuum by introducing small delays during absorption and subsequent re-emission of the photon, thus reducing the <u>average</u> propagation speed of the photons in the vacuum (compared to the light speed of photons between absorption/re-emission events).***

Our examination of the physics literature has not revealed any previous work on a quantum time delay analysis of photon propagation through the quantum vacuum, presumably because of the precedent set by Einstein's postulate of light speed constancy in the vacuum under all circumstances. We will take the position that the delays due to photon scattering through the quantum vacuum are real. These delays reduce the much faster and absolutely fixed 'low-level light velocity $c_l$' (defined as the photon velocity between vacuum particle scattering events) to the average observed light velocity 'c' in the vacuum (300,000 km/sec) that we observe in our actual experiments.

Furthermore, we propose that the quantum vacuum introduces a sort of Vacuum Index of Refraction '$n_{vac}$' (compared to a vacuum without all virtual particles) such that $c = c_l / n_{vac}$. If this is true, what is the low-level light velocity? It is unknown at this time, but it must be significantly larger than 300,000 km/sec. In fact we believe that the vacuum index of refraction '$n_{vac}$' ***must be very large*** because of the ***high density*** of virtual particles in the vacuum. This concept is required in EMQG theory, and has become central to understanding the equivalence principle and 4D space-time curvature in accelerated frames and in gravitational fields (Appendix A).

### 3.4   FIZEAU EFFECT: LIGHT VELOCITY IN A MOVING MEDIA (** Optional)

It also has been known for over a century that the velocity of light in a moving medium differs from its value in the same stationary medium. Fizeau demonstrated this experimentally in 1851 (ref. 41). For example, with a current of water (with refractive index of the medium of n=4/3) flowing with a velocity V of about 5 m/sec, the relative variation in the light velocity is $10^{-8}$ (which he measured by use of interferometry). Fresnel first derived the formula (ref. 41) in 1810 with his ether dragging theory. The resulting formula relates the longitudinal light velocity '$v_c$' moving in the same direction as a transparent medium of an index of refraction 'n' defined such that 'c/n' is the light velocity in the stationary medium, which is moving with velocity 'V' (with respect to the laboratory frame), where c is the velocity of light in the vacuum:

Fresnel Formula:   $v_c = c/n + (1 - 1/n^2) V$ (3.41)

Why does the velocity of light vary in a moving (and non-moving) transparent medium? According to the principles of special relativity, the velocity of light is a constant in the vacuum with respect to all inertial observers. When Einstein proposed this postulate, he was not aware that the vacuum is not empty. However he was aware of Fresnel's formula and derived it by the special relativistic velocity addition formula for parallel velocities (to first order). According to special relativity, the velocity of light relative to the proper frame of the transparent medium depends only on the medium. The velocity of light in the stationary medium is defined as 'c/n'. Recall that velocities u and v add according to the formula: $(u + v) / (1 + uv/c^2)$
Therefore:

$$v_c = [c/n + V] / [1 + (c/n)(V)/c^2] = (c/n + V) / (1 + V/(nc)) \approx c/n + (1 - 1/n^2) V$$
(3.42)

The special relativistic approach to deriving the Fresnel formula does not say much about the actual quantum processes going on at the atomic level. At this scale, there are several explanations for the detailed scattering process in conventional physics. We investigate these different approaches in more detail below.

3.5     LORENTZ SEMI-CLASSICAL PHOTON SCATTERING (** Optional)

The microscopic theory of the light propagation in matter was developed as a consequence of Lorentz's non-relativistic, semi-classical electromagnetic theory. We will review and summarize this approach to photon scattering, which will not only prove useful for our analysis of the Fizeau effect, but has become the basis of the 'Fizeau-like' scattering of photons in the accelerated quantum vacuum near large gravitational fields (EMQG theory, Appendix A).

To understand what happens in photon scattering inside a moving medium, imagine a simplified one-dimensional quantum model of the propagation of light in a refractive medium. The medium consisting of an idealized moving crystal of velocity 'V', which is composed of evenly spaced, point-like atoms of spacing 'l'. When a photon traveling between atoms at a speed 'c' (vacuum light speed) encounters an atom, that atom absorbs it and another photon of the same wavelength is emitted after a time lag 'τ'. In the classical wave interpretation, the scattered photon is out of phase with the incident photon. We can thus consider the propagation of the photon through the crystal is a composite signal. As the photon propagates, part of the time it exists in the atom (technically, existing as an electron bound elastically to some atom), and part of the time as a photon propagating with the undisturbed low-level light velocity 'c'. When the photon changes existence to being a bound electron, the velocity is 'V'. From this, it can be shown (ref. 41, an exercise in algebra and geometry) that the velocity of the composite signal '$v_c$' (ignoring atom recoil, which is shown to be negligible) is:

$$v_c = c \ [1 + (V\tau/l) (1 - V/c)] / [1 + (c\tau/l) (1 - V/c)] \tag{3.51}$$

If we set V=0, then $v_c = c / (1 + c\tau/l) = c/n$. Therefore, $\tau/l = (n – 1)/c$. Inserting this in the above equations give:

$$v_c = [(c/n) + (1 – 1/n) V (1 - V/c)] /[1 - (1 – 1/n)(V/c)] \approx c/n + (1 – 1/n^2) V$$
(to first order in V/c). (3.52)

Again, this is Fresnel's formula. Thus the simplified non-relativistic atomic model of the propagation of light through matter explains the Fresnel formula to the first order in V/c through the simple introduction of a scattering delay between photon absorption and subsequent re-emission. This analysis is based on a semi-classical approach. What does quantum theory say about this scattering process? The best theory we have to answer this question is QED.

3.6     PHOTON SCATTERING IN THE ACCELERATED VACUUM (** **Optional**)

Anyone who believes in the existence of virtual fermion particles in the quantum vacuum that carry mass, will acknowledge the existence of a coordinated general downward acceleration of these virtual particles near any large gravitational field. In EMQG (Appendix A) gravitons from the real fermions on the earth exchange gravitons with the virtual fermions of the vacuum (which carry electric charge), causing a downward acceleration. The virtual particles of the quantum vacuum (now accelerated by a large mass) acts on light (and matter) in a similar manner as a stream of moving water acts on light in the Fizeau effect. How does this work mathematically?

Again, it is impossible to compute the interaction of an accelerated collection of virtual particles of the quantum vacuum with light exactly. However, a simplified model can yield useful results. We will proceed using the semi-classical model proposed by Lorentz, above. We have defined the raw light velocity '$c_r$' (EMQG, ref. 1) as the photon velocity in between virtual particle scattering. Recall that raw light velocity is the shifting of the photon information pattern by one cell at every clock cycle on the CA, so that in fundamental units it is an absolute constant. Again, we assume that the photon delay between absorption and subsequent re-emission by a virtual particle is '$\tau$', and the average distance between virtual particle scattering is 'l'. The scattered light velocity $v_c(t)$ is now a function of time, because we assume that it is constantly varying as it moves downwards towards the surface in the same direction of the virtual particles. The virtual particles move according to: a =gt, where $g = GM/R^2$.

Therefore we can write the velocity of light after scattering with the accelerated quantum vacuum:

$$v_c(t) = c_r \ [1 + (gt\tau/l) \ (1 - gt/c_r)] \ / \ [1 + (c_r\tau/l) \ (1 - gt/c_r)] \tag{3.61}$$

If we set the acceleration to zero, or $gt = 0$, then $v_c(t) = c_r \ / \ (1 + c_r\tau/l) = c_r/n$. Therefore, $\tau/l = (n - 1)/c_r$. Inserting this in the above equation gives:

$$v_c(t) = [(c_r/n) + (1 - 1/n) \ gt \ (1 - gt/c_r)] \ / \ [1 - (1 - 1/n)(gt/c_r)] \approx c_r/n + (1 - 1/n^2) \ gt$$
to first order in $gt/c_r$. (3.62)

Since the average distance between virtual charged particles is very small, the photons (which are always created at velocity $c_r$) spend most of the time existing as some virtual charged particle undergoing downward acceleration. Because the electrically charged virtual particles of the quantum vacuum are falling in their brief existence, the photon *effectively* takes on the <u>same downward acceleration</u> as the virtual vacuum particles (as an average over macroscopic distances). In other words, because the index of refraction of the quantum vacuum 'n' is so large (relative to no vacuum particles), and because $c = c_r/n$ and we can write in equation 3.62:

$$v_c(t) = c_r/n + (1 - 1/n^2) \ gt = c + gt = c \ (1 + gt/c) \ \text{if} \ n \gg 1. \tag{3.63}$$

Similarly, for photons going against the flow (upwards): $v_c(t) = c \ (1 - gt/c)$ (3.64)

This formula is used in EMQG for the variation of light velocity near a large gravitational field, and leads to the correct amount of general relativistic space-time curvature taking into account some additional assumptions as shown in Appendix A. It is this path Einstein followed

## 4. NON-LOCALITY AND SUPERLUMINAL PHOTONIC TUNNELING

Is there any other evidence in physics for phenomena that potentially exhibit faster-than-light propagation? In quantum mechanics there definitely exist such phenomena:

(1) **The Quantum Non-Locality of quantum entangled particles, which *apparently* communicate each other's quantum state faster-than-light**.
(2) ***Apparent* faster-than-light tunneling of photons through a potential barrier, which is classically too large for the photon to penetrate.**

Both phenomena are well known and described in standard text books on quantum theory. A good account of both is given in an article titled "FASTER THAN LIGHT?". In Scientific American, August 1993 by R. Chiao, P. Kwiat, A. Steinberg (ref. 26). Non-locality is an effect where two particles (or more) are causally connected or entangled (in other words where the two particle's wave functions are dependent on each other), can influence each other apparently instantaneously no matter how far apart they are. A famous example of entanglement is the famous Einstein EPR proposal of 1935 (ref. 10), and subsequent experimental verification by Aspect (ref. 12). It was J.S. Bell (ref. 28) that first derived a set of inequalities that Nature should obey if locality and reality were obeyed, and in which are violated by quantum mechanics. In some interpretations of quantum theory this appears to be contradiction of strict Einstein locality or causality, and therefore provides evidence for faster-than-light signaling. However to our knowledge no one has been able to devise a method of sending information faster than light from one location to another using quantum non-locality methods (excluding the claims by Nimtz, ref. 27.

Quantum Tunneling is a phenomenon that is in some sense related to non-locality. In photon tunneling, a photon has a finite probability of moving through a barrier that it should not be able to pass through according to classical physics. What is remarkable about tunneling of photons (or for other types of quantum particles) is that when a measurement is made for the tunneling velocity, one finds that it is greater than the velocity of light in a vacuum. However some physicists maintain that it is not possible to talk about the photon actually having a definite velocity while it passes through the barrier. Indeed the act of assigning a definite time to the tunneling process has also been questioned (ref. 24). These arguments appeal to the probabilistic nature of the wave packet that describes the photon. The problem of defining the tunneling time of photon penetration through a barrier has a long history, which dates back to the 1920's, when Hund first proposed the quantum mechanical "barrier penetration" phenomena. Quantum tunneling has important applications in electronics, and the very operation of the tunnel diode depends on the existence of the quantum tunneling phenomenon.

An excellent account of quantum tunneling of photons is given by R. Y. Chiao, P. Kwiat, and A. Steinberg in the August 1993 (ref. 26) Scientific American magazine titled "Faster than Light?". These authors claim that their "***Experiments in quantum optics show that two distant events can influence each other faster than any signal could have traveled between them***." They report that during several days of data collection (of more than one million photons tunneling through their barrier), that on average the tunneling photons arrived before the unimpeded photons. Their results imply that the average tunneling velocity for the photons is about 1.7 times that of light (ref. 25). What is even more bizarre is that the 'tunneling velocity' (which is a questionable concept) does not depend on the width of the barrier!

R. Chiao et al. provide an explanation for faster-than-light tunneling that is not based on the concept of a tunneling velocity or a tunneling time. They point out that the photon's quantum mechanical wave function of the tunneling photon is greatly reduced in amplitude in comparison to the unimpeded photon's wave function. Recall that the amplitude of the wave function at a point represents the probability of finding a photon at that point. They point out that the center of the photon wave packet is the place of greatest probability of detection in the experiment. They claim that the wave front of the tunneling and unimpeded photons move together at the same rate, but that the tunneling photons arrive first because of the change in wave shape. They illustrate this elegantly with racing tortoises, where the two noses of the tortoises are locked in step, but the smaller one has a narrower wave packet width and subsequently is detected first (illustration on bottom of page 55, ref. 26).

Other experiments by G. Nimitz using microwave frequency photons instead of light, demonstrate that the microwave 'tunneling velocity' is 4.7 times light speed. Furthermore in order to illustrate that tunneling can convey information faster-than-light, they modulated the microwave beam with the audio track of the 1$^{st}$ movement of Mozart's 40$^{th}$ symphony. They report that they were able to send this message through the barrier at 4.7 times light speed (ref. 27).

The proposed LVC experiments borrow much of the same techniques used by R. Chiao et. al. to measure the increase in light velocity in quantum tunneling. The task is similar, that is to compare the arrival times of photons from two different photon paths that started at the same time, where the difference in arrival times is incredibly small and hard to detect. The speed of light is so great at laboratory distances that conventional electronics is tens of thousands of times too slow to measure the small differences in light arrival time. To solve this problem they used twin photon interferometers to measure the required time delays, a technique that we will borrow for our proposed experiments.

## 5. THE PROPOSED CASIMIR LIGHT VELOCITY EXPERIMENTS

We propose experiments to look for the Light Velocity Casimir (LVC) effect and to measure the Casimir vacuum index of refraction. Figure 1 shows the conceptual block diagram of the first experiment to observe the increase in light velocity. Figure 3 gives a more detailed account, which we will describe later. Here we have two identical light paths that originated from a common light source such as a laser, where one light path travels straight through the vacuum unimpeded, and becomes our standard reference path for the light velocity in the vacuum. The other light path travels between two transparent electrically conducting, Casimir plates in a vacuum, which are closely spaced and have an adjustable plate spacing 'd'. When the laser light source is switched on, the light path through the Casimir plates arrives at its detector first, where the arrival time becomes *sooner* with *decreasing* plate spacing.

*NOTE: Light must propagate perpendicular through the plates in the LVC experiment, because it is in this direction that vacuum density decreases. If light travels parallel through the Casimir plates (which would offer a longer path to increase light velocity), the vacuum density and light velocity along this direction are **not changed**. This illustrates that the vacuum process inside the Casimir plates is a dynamic effect. If the Casimir plates are visualized as being part of a rectangular enclosure and air is pumped out of the box, the velocity of light would increase compared to the velocity of light in air, no matter what direction light traveled in the box. If the two ends of the rectangular box are removed, then the inside and outside air pressure would balance. This is not so for the Casimir plates. Even though the two ends of the 'Casimir' box are removed, the Casimir plates maintain a decreased vacuum density, but only in the direction*

*perpendicular to the plates. This pressure change is maintained dynamically by the quantum vacuum process discussed in section 2.1.*

We claim that the front velocity of the light traveling through the Casimir plates will exceed the velocity of light in the ordinary vacuum, as defined in section 1.1. In order to enhance the magnitude of the light velocity increase, it is desirable to increase the optical path length for the laser light through the Casimir plates. This can be done by optically arranging a set of mirrors to direct the light back and forth several times through the Casimir plates (which must also be done in the reference path). In order to simplify the discussions, this arrangement is not shown in any of the diagrams, and the method used depends on the detailed experimental arrangement chosen. The two beams are then routed to two individual detectors (Figure 3). The detectors electronic outputs are fed to an electronic instrument that can accurately measure time delays between the two outputs.

Although we have been unable to compute the value of the light velocity increase, we expect it to be very small. We believe that using an ordinary laser beam (which contains enormous numbers of photons) as a light source will not be effective in measuring the LVC effect. Instead we borrow some techniques from quantum optics and from the measurement of optical tunneling times to detect the LVC effect for single photons. The time that it takes light to transverse the Casimir plate spacing (assuming the plates are not there) with a 1μm spacing is 33 fs ($33 \times 10^{-15}$ seconds)! Therefore we expect that the time to propagate through the plates to be smaller than this value. That means we would like to have perhaps 0.1 fs resolution in time. Currently the best photon detectors only have a picosecond-scale response time, which is not fast enough for this application. However a device called the 'Hong-Ou-Mandel Interferometer' has femtosecond-scale time resolution, which is ideal for this experiment.

Figure 3 shows a conceptual Hong-Ou-Mandel interferometer-based setup that should be capable of measuring the Casimir light velocity increase for individual photons! It is very similar to the experimental arrangement to measure the quantum tunneling of individual photons through a barrier used by R. Chiao et al. (ref. 25 and 26). In order to make sure both photons start traveling at the same time through the apparatus we suggest the use of a 'Spontaneous Parametric Down-Conversion' crystal, which absorbs the incoming ultraviolet photon from an argon laser and emits two new photons simultaneously, and which are strongly correlated. The energies of the two photons equal the energy of the incoming photon. Furthermore the two photons are quantum mechanically entangled, which is beneficial to overcoming potential sources of errors in performing the experiment (ref. 25). The two photons reflect off the two mirrors and travel through two equal length paths and meet at the beam splitter. One path is the reference vacuum path, and the other is the Casimir plate path, where the plates are originally removed to calibrate the optical paths to null the interferometer (ref. 25).

The advantage of using the Hong-Ou-Mandel Interferometer is that it results in a narrow null in the coincidence count rate as a function of the relative delay between the two photons, a destructive interference effect that was first observed by Hong, Ou, and Mandel. The narrowness of the coincidence minimum combined with a good signal to noise ratio should provide a measurement of the relative delay between the two photons to a precision of ±0.2 fs (ref. 25).

There are four possible outcomes at the beam splitter: both photons might pass through, both might be reflected from the beam splitter, both might go off to one side (one reflected, and one transmitted), or both may go off the other side. We are interested in the first two cases, where the two photons reach different detectors that result in ***coincidence detection***. We adjust the optical lengths until coincidence detection disappears. This means that any deviation in the relative velocities of the two path's results will become *readily apparent* because of the narrowness of the interference null. The detectors chosen for the tunneling experiment (ref. 25) are Geiger-mode silicon avalanche photo-diodes. Aslo, care must be taken not to choose a Casimir plate spacing which nulls out the photon (multiples of the photon frequency), since the Casimir plates are electrically conductive.

The Vacuum Casimir Index of Refraction given by $n_{vac} = c / c_c$ (which should be slightly less than one) where '$c_c$' is the light velocity between the Casimir plates. This can be measured in the experimental arrangement of figure 3, assuming the LVC effect is observed. In order to calculate $n_{vac}$ one must measure $c_c$. Let 'h' be the optical path length of the reference leg of the interferometer. Let 'd' be the distance between the Casimir plates. Let '$\delta t$' be the time a photon takes to traverse the Casimir plates. Let '$\Delta t$' be the time a photon takes to traverse the plate spacing, with no plates present. The interferometer measures the time $t_{diff} = \Delta t - \delta t$, or the decrease in the relative time of propagation of the photon through Casimir plate distance 'd'. In order to calculate $n_{vac}$:

- Determine the optical length 'h' through the reference leg of the interferometer, which is also equal to the optical length of the Casimir plate leg of the interferometer.
- Determine the time delay $t_{diff} = \Delta t - \delta t$, which is also the time difference between the arrival of a photon first at detector 2 followed by the detection at detector 1.
- Calculate $\Delta t = d/c$, and then calculate $\delta t = \Delta t - t_{diff}$.
- It follows that the light velocity inside the Casimir plates is given by: $c_c = d / \delta t$, which should be slightly less than 'c' in the reference leg.
- Finally we have the Vacuum Casimir Index of Refraction:   $n_{vac} = c / c_c$

Figure 2 shows an alternate way to demonstrate the Light Velocity Casimir effect, based on the classical idea of light refraction. Conceptually we start with a single laser light source and direct the light at a shallow angle '$t_0$' that is close to the perpendicular of the two transparent, electrically conducting Casimir plates (see the bubble in figure 2). This must be so, because the refraction only happens in the nearly perpendicular direction of the Casimir plates (see the note at the beginning of section 5).

The light deflects at the first plate by an angle $t_1$, such that $n_{vac} = \sin(t_0) / \sin(t_1)$. When the light exits the second plate it returns to the original direction because it goes to the normal vacuum state which is an increase in density. The end result is that the light beam is still in line with the incoming beam, but slightly shifted to the right. In these discussions we ignore any refraction through the transparent plate material, which must be taken into account when performing the actual experiment. Because the spacing between the plates is quite small, the light bending effect would be very hard to detect with only a two Casimir plate experiment.

In order to enhance the bending effect it is desirable to have a series of transparent and conductive Casimir plates, such that the spacing between the subsequent plates decreases with distance. Figure 2 shows an example arrangement of four such Casimir plates for illustration purposes. With this arrangement the exit angle of the light is permanently different from the original angle, because the vacuum density between the plates varies in different steps between each new pair of plates. This means that over a large distance the angle can be easily measured. In practice the resulting angle is going to be extremely small, because of the very slight differences in energy-density per pair of plates.

However a small angle can translate to a large shift or deflection in the total light path, when the light is allowed to travel over a large distance in normal air. For example if the difference in angle of the incoming and outgoing light paths is $\Delta\theta = 0.001$ degrees (which is only 3.6 seconds of arc), at a distance of h=100 meters the difference in position of the expected light beam is given by: h tan($\Delta\theta$) = 1.7 millimeters, which is a distance that can be easily measured.

## 6.     CONCLUSIONS

1. We believe that the velocity of light (specifically the front velocity) propagating in vacuum inside (and perpendicular) two closely spaced, electrically conducting and transparent plates called the Casimir plates, will *increase* inside the plates compared to the light velocity in the normal vacuum, measured in the laboratory frame. This conclusion should be subjected to future experimental verification. We have proposed two such experiments; an interferometer, which is designed to resolve the difference in arrival time of individual photons, and another experiment to measure the refraction of light propagating at a shallow angle (nearly perpendicular) through a series of Casimir plates with gradually decreasing plate separation.
2. We believe that there must exist a 'Vacuum Casimir Index of Refraction' called '$n_{vac}$' for light traveling from outside, and through the Casimir plates in vacuum. The Casimir vacuum index of refraction is defined as the ratio of the velocity of light in normal vacuum conditions divided by the light velocity measured propagating inside the Casimir plates in vacuum '$c_c$'. The vacuum Casimir index of refraction '$n_{vac}$' is thus defined as: $n_{vac} = c / c_c$, which should be slightly less than one. This should manifest itself as a slight change in direction of the light beam through the Casimir plates in accordance with Snell's law of refraction: $n = \sin\theta_0 / \sin\theta_1$.

## 8.    FIGURE CAPTIONS

The captions for the figures are shown below:

Fig. 1 - Schematic Diagram of the Light Velocity Casimir Effect
Fig. 2 - Experiment to measure the Refraction of Light through Multiple Casimir Plates
Fig. 3 - An Interferometer Setup for the Light Velocity Casimir Experiment

# THE LIGHT VELOCITY CASIMIR EFFECT EXPERIMENTS

### FIGURE #1 - Schematic Diagram of the Light Velocity Casimir Effect

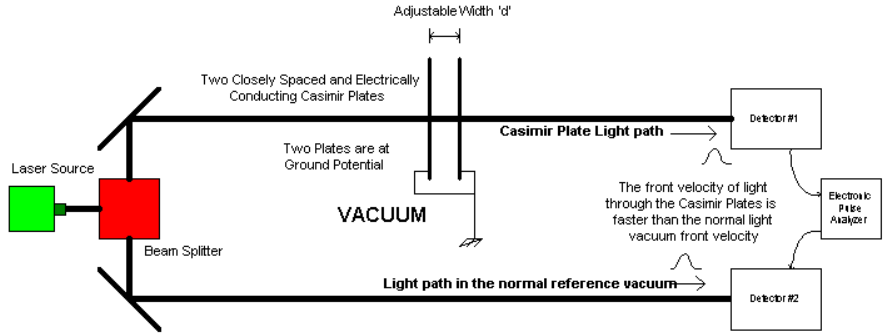

### FIGURE #2 - Experiment to measure the refraction of Light through multiple Casimir Plates with the plate spacing gradually decreasing

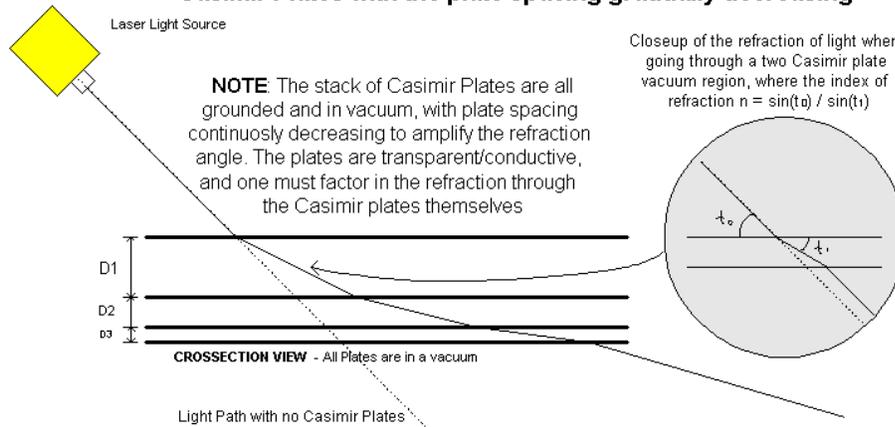

NOTE: By varying the spacing of the Multiple Casimir plates the refraction angle of the light is increased and permanently changed as light leaves the plates. The total refraction angle can be easily observed over a large distance and can be used to calculate the Casimir index of refraction of light in the vacuum.

## Figure #3 - AN INTERFEROMETER SETUP FOR THE LIGHT VELOCITY CASIMIR EXPERIMENT

A sensitive interferometer based arrangement of the Light Velocity Casimir Experiment. The interferometer is calibrated with the Casimir plate spacing being very large, and the two optical paths adjusted so that there is a null reading from the detectors. As the plates are moved closer together, Detector 2 reads pulses before Detector 1

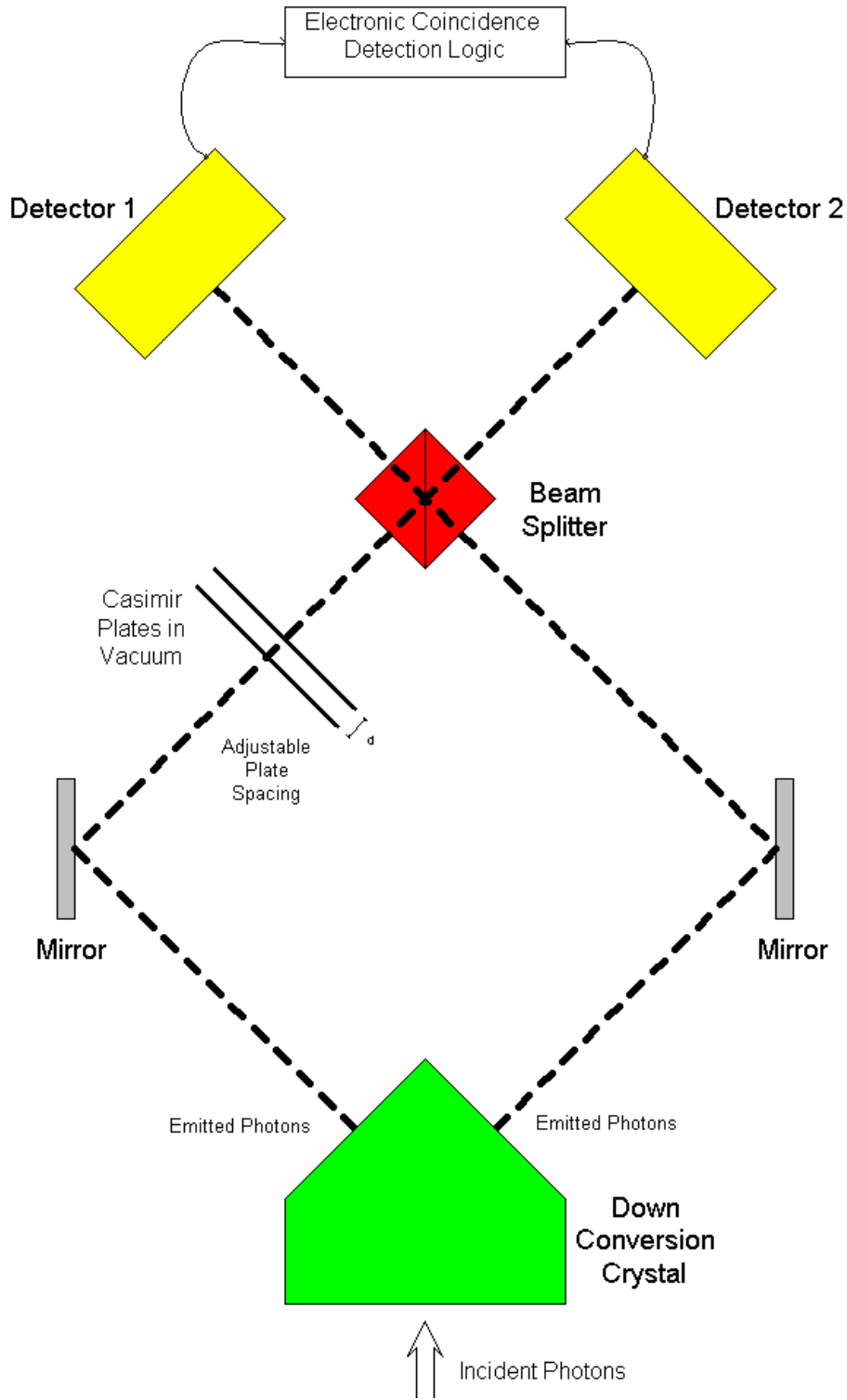

34# APPENDIX A: BRIEF REVIEW OF EMQG

This appendix gives a very brief review of Electromagnetic Quantum Gravity (EMQG) and it's connection to the quantum vacuum and the Casimir Light Velocity effect. The full paper can be found in reference A1. This review is intended to briefly summarize the essential ideas of EMQG and the central role that the quantum vacuum plays in EMQG.

We have developed a new approach to the unification of quantum theory with general relativity referred to as Electro-Magnetic Quantum Gravity or EMQG (ref. 1). Figure A1 at the end of the appendix illustrates the relationship between EMQG and the rest of physical theory. EMQG has its origins in Cellular Automata (CA) theory (ref. 2,4,9 and 34), and is also based on the new theory of inertia that has been proposed by R. Haisch, A. Rueda, and H. Puthoff (ref. 5) known here as the HRP Inertia theory. These authors suggest that inertia is due to the strictly local force interactions of charged matter particles with their immediate background virtual particles of the quantum vacuum. They found that inertia is caused by the magnetic component of the Lorentz force, which arises between what the author's call the charged 'parton' particles in an accelerated reference frame interacting with the background quantum vacuum virtual particles. The sum of all these tiny forces in this process is the source of the resistance force opposing accelerated motion in Newton's F=MA. We have found it necessary to make a small modification of HRP Inertia theory as a result of our investigation of the principle of equivalence. The modified version of HRP inertia is called "Quantum Inertia" (or QI). In EMQG, a new elementary particle is required to fully understand inertia, gravitation, and the principle of equivalence (described in the next section). This theory also resolves the long outstanding problems and paradoxes of accelerated motion introduced by Mach's principle, by suggesting that the vacuum particles themselves serve as Mach's universal reference frame (for <u>acceleration</u> only), without violating the principle of relativity of constant velocity motion. In other words, our universe offers no observable reference frame to gauge inertial frames, because the quantum vacuum offers no means to determine absolute constant velocity motion. However for accelerated motion, the quantum vacuum plays a very important role by offering a resistance to acceleration, which results in an inertial force opposing the acceleration of the mass. Thus the very existence of inertial force reveals the absolute value of the acceleration with respect to the net statistical average acceleration of the virtual particles of the quantum vacuum. Reference 14 offers an excellent introduction to the motion of matter in the quantum vacuum, and on the history of the discovery of the virtual particles of the quantum vacuum.

### (A-1) EMQG and the Quantum Theory of Inertia

EMQG theory presents a unified approach to Inertia, Gravity, the Principle of Equivalence, space-time Curvature, Gravitational Waves, and Mach's Principle. These apparently different phenomena are the common results of the quantum interactions of the real (charged) matter particles (of a mass) with the surrounding virtual particles of the quantum vacuum through the exchange of two force particles: the photon and the graviton. Furthermore, the problem of the cosmological constant is solved automatically in the framework of EMQG. This new approach to quantum gravity is definitely *non-geometric* on the tiniest of distance scales (Plank Scales of distance and time). This is because the large scale relativistic 4D space-time curvature is caused purely by the accelerated state of virtual particles of the quantum vacuum with respect to a mass, and their discrete interactions with real matter particles of a mass through the particle force exchange process. Because of this departure from a universe with fundamentally curved space-time, EMQG is a complete change in paradigm over conventional gravitational physics. This paper should be considered as a framework, or outline of a new approach to gravitational physics that will hopefully lead to a full theory of quantum gravity.

We modified the HRP theory of Inertia (ref. 5) based on a detailed examination of the principle of equivalence. In EMQG, the modified version of inertia is known as "Quantum Inertia", or QI. In EMQG, a new elementary particle is required to fully understand inertia, gravitation, and the principle of equivalence. All matter, including electrons and quarks, must be made of nature's most fundamental mass unit or particle, which we call the 'masseon' particle. These particles contain one fixed, fundamental 'quanta' of both inertial and gravitational mass. The masseons also carry one basic, smallest unit or quanta of electrical charge as well, of which they can be either positive or negative. Masseons exist in the particle or in the anti-particle form (called anti-masseon), that can appear at random in the vacuum as virtual masseon/anti-masseon particle pairs of opposite electric charge and opposite 'mass charge'. The earth consists of ordinary masseons (with no anti-masseons), of which there are equal numbers of positive and negative electric charge varieties. In HRP Inertia theory, the electrically charged 'parton' particles (that make up an inertial mass in an accelerated reference frame) interact with the background vacuum electromagnetic zero-point-field (or virtual photons) creating a resistance to acceleration called inertia. We have modified this slightly by postulating that the real masseons (that make up an accelerating mass) interacts with the surrounding, virtual masseons of the quantum vacuum, electromagnetically (although the details of this process are still not fully understood). The properties of the masseon particle and gravitons are developed later.

**(A-2) EMQG and the Quantum Origin of Newton's Laws of Motion**

We are now in a position to understand the quantum nature of Newton's classical laws of motion. According to the standard textbooks of physics, Newton's three laws of laws of motion are:

An object at rest will remain at rest and an object in motion will continue in motion with a constant velocity unless it experiences a net external force.
The acceleration of an object is directly proportional to the resultant force acting on it and inversely proportional to its mass. Mathematically: $\Sigma F = ma$, where F and a are vectors.

If two bodies interact, the force exerted on body 1 by body 2 is equal to and opposite the force exerted on body 2 by body 1. Mathematically: $F_{12} = -F_{21}$.

Newton's first law explains what happens to a mass when the resultant of all external forces on it is zero. Newton's second law explains what happens to a mass when there is a nonzero resultant force acting on it. Newton's third law tells us that forces always come in pairs. In other words, a single isolated force cannot exist. The force that body 1 exerts on body 2 is called the action force, and the force of body 2 on body 1 is called the reaction force.

In the framework of EMQG theory, Newton's first two laws are the direct consequence of the (electromagnetic) force interaction of the (charged) elementary particles of the mass interacting with the (charged) virtual particles of the quantum vacuum. Newton's third law of motion is the direct consequence of the fact that all forces are the end result of a boson particle exchange process.

### (A-3) NEWTON'S FIRST LAW OF MOTION:

In EMQG, the first law is a trivial result, which follows directly from the quantum principle of inertia (postulate #3, appendix A-11). First a mass is at relative rest with respect to an observer in deep space. If no external forces act on the mass, the (charged) elementary particles that make up the mass maintain a *net acceleration* of zero with respect to the (charged) virtual particles of the quantum vacuum through the electromagnetic force exchange process. This means that no change in velocity is possible (zero acceleration) and the mass remains at rest. Secondly, a mass has some given constant velocity with respect to an observer in deep space. If no external forces act on the mass, the (charged) elementary particles that make up the mass also maintain a *net acceleration* of zero with respect to the (charged) virtual particles of the quantum vacuum through the electromagnetic force exchange process. Again, no change in velocity is possible (zero acceleration) and the mass remains at the same constant velocity.

### (A-4) NEWTON'S SECOND LAW OF MOTION:

In EMQG, the second law is the quantum theory of inertia discussed above. Basically the state of *relative* acceleration of the charged virtual particles of the quantum vacuum with respect to the charged particles of the mass is what is responsible for the inertial force. By this we mean that it is the tiny (electromagnetic) force contributed by each mass particle undergoing an acceleration 'A', with respect to the net statistical average of the virtual particles of the quantum vacuum, that results in the property of inertia possessed by all masses. The sum of all these tiny (electromagnetic) forces contributed from each charged particle of the mass (from the vacuum) is the source of the total inertial resistance force opposing accelerated motion in Newton's F=MA. Therefore, inertial mass 'M' of a mass simply represents the total resistance to acceleration of all the mass particles.

### (A-5) NEWTON'S THIRD LAW OF MOTION:

According to the boson force particle exchange paradigm (originated from QED) all forces (including gravity, as we shall see) result from particle exchanges. Therefore, the force that body 1 exerts on body 2 (called the action force), is the result of the emission of force exchange particles from (the charged particles that make up) body 1, which are readily absorbed by (the charged particles that make up) body 2, resulting in a force acting on body 2. Similarly, the force of body 2 on body 1 (called the reaction force), is the result of the absorption of force exchange particles that are originating from (the charged particles that make up) body 2, and received by (the charged particles that make up) body 1, resulting in a force acting on body 1. An important property of charge is the ability to readily emit <u>and</u> absorb boson force exchange particles. Therefore, body 1 is both an emitter and an absorber of the force exchange particles. Similarly, body 2 is also both an emitter and an absorber of the force exchange particles. This is the reason that there is both an action and reaction force. For example, the contact forces (the mechanical forces that Newton was thinking of when he formulated this law) that results from a person pushing on a mass (and the reaction force from the mass pushing on the person) is really the exchange of photon particles from the charged electrons bound to the atoms of the person's hand and the charged electrons bound to the atoms of the mass on the quantum level. Therefore, on the quantum level there is really is no contact here. The hand gets very close to the mass, but does not actually touch. The electrons in one's hand exchange photons with the electrons in the mass. The force exchange process works both directions in equal numbers, because all the electrons in the hand and in the mass are electrically charged and therefore the exchange process gives forces that are equal and opposite in both directions.

### **(A-6) Introduction to the Principle of Equivalence and EMQG**

Are virtual particle force exchange processes originating from the quantum vacuum also present for gravitational mass? The answer turns out to be a resounding yes! As we suggested, there is some evidence of the interplay between the virtual particles of the quantum vacuum and gravitational phenomena. In order to see how this impacts our understanding of the nature of gravitational mass, we found it necessary to perform a thorough investigation of Einstein's Principle of Equivalence of inertial and gravitational mass in general relativity under the guidance of the new theory of quantum inertia.

We have uncovered some theoretical evidence that the SEP may not hold for certain experiments. There are two basic theoretical problems with the SEP in regard to quantum gravity. First, if gravitons (the proposed force exchange particle) can be detected with some new form of a sensitive graviton detector, we would be able to distinguish between an accelerated reference frame and a gravitational frame with this detector. This is because accelerated frames would have virtually no graviton particles present, whereas a large gravitational field like the earth has enormous numbers of graviton particles associated with it. Secondly, theoretical considerations from several authors regarding the emission of electromagnetic waves from a uniformly accelerated charge, and the lack of radiation from the same charge subjected to a static gravitational field leads us to question the validity of the SEP for charged particles radiating electromagnetically.

How does the WEP hold out in EMQG? The WEP has been tested to a phenomenal accuracy (ref 24.) in recent times. Yet in our current understanding of the WEP, we can only specify the accuracy as to which the two different mass values (or types) have been shown experimentally to be equal inside an inertial or gravitational field. There exists no physical or mathematical proof that the WEP is precisely true. It is still only a postulate of general relativity. We have applied the recent work on quantum inertia (ref. 5) to the investigation of the weak principle of equivalence, and have found theoretical reasons to believe that the WEP is not precisely correct when measured in extremely accurate experiments. Imagine an experiment with two masses; one mass $M_1$ being very large in value, and the other mass $M_2$ is very small ($M_1 >> M_2$). These two masses are dropped simultaneously in a uniform gravitational field of 1g from a height 'h', and the same pair of masses are also dropped inside a rocket accelerating at 1g, at the same height 'h'. We predict that there should be a minute deviation in arrival times on the surface of the earth (only) for the two masses, with the heavier mass arriving just slightly ahead of the smaller mass. This is due to a small deviation in the magnitude of the force of gravity on the mass pair (in favor of $M_1$) on the order of $(N_1-N_2)i * \delta$, where $(N_1-N_2)$ is the difference in the low level mass specified in terms of the difference in the number of masseon particles in the two masses (defined latter) times the single masseon mass 'i', and $\delta$ is the ratio of the gravitational to electromagnetic forces for a single (charged) masseon. This experiment is very difficult to perform on the earth, because $\delta$ is extremely small ($\approx 10^{-40}$), and $\Delta N = (N_1-N_2)$ cannot in practice be made sufficiently large in order to produce a measurable effect. However, inside the accelerated rocket, the arrival times are <u>exactly</u> identical for the same pair of masses. This, of course, violates the principle of equivalence, since the motion of the masses in the inertial frame is slightly different then in the gravitational frame. This imbalance is minute because of the dominance of the strong electromagnetic force, which is also acting on the masseons of the two masses from the virtual particles of the quantum vacuum. This acts to stabilize the fall rate, giving us nearly perfect equivalence.

This conclusion is based on the discovery that the weak principle of equivalence results from lower level physical processes. Mass equivalence arises from the equivalence of the force generated between the net statistical average acceleration vectors of the charged matter particles inside a mass with respect to the immediate surrounding quantum vacuum virtual particles inside an accelerating rocket. This is almost exactly the same physical force occurring between the stationary (charged) matter particles and the immediate surrounding accelerating virtual particles of the same mass near the earth. It turns out that equivalence is not perfect in the presence of a large gravitational field like the earth. Equivalence breaks down due to an extremely minute force imbalance in favor of a larger mass dropped simultaneously with respect to a smaller mass. This force imbalance can be traced to the pure graviton exchange force component occurring in the gravitational field that is not present in the case of the identically dropped masses in an accelerated rocket. This imbalance contributes a minute amount of extra force for the larger mass compared to the smaller mass (due to many more gravitons exchanged between the larger mass as compared to the smaller mass), which might be detected in highly accurate measurements. In the case of the rocket, the equivalence of two different falling masses is perfect, since it is the floor of the rocket that accelerates up to meet the two masses simultaneously. Of course, the breakdown of the WEP also means the downfall of the SEP.

In EMQG, the gravitational interactions involve the same electromagnetic force interaction as found in inertia based on our QI theory. We also found that the weak principle of equivalence itself is a physical phenomenon originating from the hidden lower level quantum processes involving the quantum vacuum particles, graviton exchange particles, and photon exchange particles. In other words, gravitation is purely a quantum force particle exchange process, and is not based on low level fundamental 4D curved space-time geometry of the universe as believed in general relativity. The perceived 4D curvature is a manifestation of the dynamic state of the falling virtual particles of the quantum vacuum in accelerated frames, and gravitational frames. The only difference between the inertial and gravitation force is that gravity also involves graviton exchanges (between the earth and the quantum vacuum virtual particles, which become accelerated downwards), whereas inertia does not. Gravitons have been proposed in the past as the exchange particle for gravitational interactions in a quantum field theory of gravity without much success. The reason for the lack of success is that graviton exchange is not the only exchange process occurring in large-scale gravitational interactions; photon exchanges are also involved! It turns out that not only are there both graviton and photon exchange processes occurring simultaneously in large scale gravitational interactions such as on the earth, but that both exchange particles are almost identical in their fundamental nature (Of course, the strength of the two forces differs greatly).

The equivalence of inertial and gravitational mass is ultimately traced down to the reversal of all the relative acceleration vectors of the charged particles of the accelerated mass <u>with respect</u> to the (net statistical) average acceleration of the quantum vacuum particles, that occurs when changing from inertial to gravitational frames. The inertial mass 'M' of an object with acceleration 'a' (in a rocket traveling in deep space, away from gravitational fields) results from the sum of all the tiny forces of the charged elementary particles that make up that mass with respect to the immediate quantum vacuum particles. This inertial force is in the opposite direction to the motion of the rocket. The (charged masseon) particles building up the mass in the rocket will have a net statistical average acceleration 'a' with respect to the local (charged masseon) virtual particles of the immediate quantum vacuum. A stationary gravitational mass resting on the earth's surface has this same quantum process occurring as for the accelerated mass, but with the acceleration vectors reversed. What we mean by this is that under gravity, it is now the virtual particles of the quantum vacuum that carries the net statistical average acceleration 'A' downward. This downward virtual particle acceleration is caused by the graviton exchanges between the earth and the mass, where the mass is not accelerated with respect to the center of mass of the earth. (Note: On an individual basis, the velocity vectors of these quantum vacuum particles actually point in all directions, and also have random amplitudes. Furthermore, random accelerations occur due to force interactions between the virtual particles themselves. This is why we refer to the statistical nature of the acceleration.) We now see that the gravitational force of a stationary mass is also the <u>same</u> sum of the tiny forces that originate for a mass undergoing accelerated motion in a gravitational field from the virtual particles of the quantum vacuum according to Newton's law 'F = MA'. In other words, the same inertial force F=MA is also found hidden inside gravitational interactions of masses! Mathematically, this fact can be seen in Newton's laws of inertia and in Newton's gravitational force law by slightly rearranging the formulas as follows:

$F_i = M_i (A_i)$ ... the inertial force $F_i$ opposes the acceleration $A_i$ of mass $M_i$ in the rocket, caused by the sum of the tiny forces from the virtual particles of the quantum vacuum.

$F_g = M_g (A_g) = M_g (GM_e/r^2)$ ... the gravitational force $F_g$ is the result of a kind of an inertial force given by '$M_g A_g$' where $A_g = GM_e/r^2$ is now due to the sum of the tiny forces from the virtual particles of the quantum vacuum (now accelerating downwards).

Since $F_i=F_g$, and since the acceleration of gravity is chosen to be the same as the inertial acceleration, where the virtual particles now have: $A_g = A_i = GM_e/r^2$ , therefore $M_i=M_g$ , or the inertia mass is equal to the gravitational mass ($M_e$ is the mass of the earth). Here, Newton's law of gravity is rearranged slightly to emphasis it's form as a kind of an 'inertial force' of the form F=MA, where the acceleration ($GM_e/r^2$) is now the net statistical average downward acceleration of the quantum vacuum virtual particles near the vicinity of the earth.

This derivation is not complete, unless we can provide a clear explanation as to why $F_i=F_g$, which we know to be true from experimental observation. In EMQG, both of these forces are understood to arise from an almost identical quantum vacuum process! For accelerated masses, inertia is the force $F_i$ caused by the sum of all the tiny electromagnetic forces from each of the accelerated charged particles inside the mass; with respect to the non-accelerating surrounding virtual particles of the quantum vacuum. Under the influence of a gravitational field, the <u>same force</u> $F_g$ exists as it does in inertia, but now the quantum vacuum particles are the ones undergoing the same acceleration $A_i$ (through graviton exchanges with the earth); the charged particles of the mass are stationary with respect to earth's center. The same force arises, but the arrows of the acceleration vectors are reversed. To elaborate on this, imagine that you are in the reference frame of a stationary mass resting on the surface with respect to earth's center. An average charged particle of this mass 'sees' the virtual particles of the quantum vacuum in the same state of acceleration, as does an average charged particle of an identical mass sitting on the floor of an accelerated rocket (1 g). In other words, the background quantum vacuum 'looks' exactly the same from both points of view (neglecting the very small imbalance caused by a very large number of gravitons interacting with the mass directly under gravity, this imbalance is swamped by the strength of the electromagnetic forces existing).

These equations and methodology illustrate equivalence in a special case: that is between an accelerated mass $M_i$ and the same <u>stationary</u> gravitational mass $M_g$. In EMQG, the weak equivalence principle of gravitational and inertial frames holds for many other scenarios such as for free falling masses, for masses that have considerable self gravity and energy (like the earth), for elementary particles, and for the propagation of light. However, equivalence is *not* perfect, and in some situations (for example, antimatter discussed in section 7.1) it simply does not hold at all!

An astute observer may question why all the virtual particles (electrons, quarks, etc., all having different masses) are accelerating downwards on the earth with the same acceleration. This definitely would be the case from the perspective of a mass being accelerated by a rocket (where the observer is accelerating). Since the masses of the different types of virtual particles are all different according to the standard model of particle physics, why are they all falling at the same rate? Since we are trying to derive the equivalence principle, we cannot invoke this principle to state that all virtual particles must be accelerating downward at the same rate. It turns out that the all quantum vacuum virtual particles are accelerating at the same rate because all particles with mass (virtual or not) are composed of combinations of a new fundamental "masseon" particle, which carries just one fixed quanta of mass. Therefore, all the elementary virtual masseon particles of the quantum vacuum are accelerated by the same amount. These masseons can bind together to form the familiar particles of the standard model, like virtual electrons, virtual positrons, virtual quarks, etc. Recalling that the masseon also carries electrical charge, we see that all the constituent masseons of the quantum vacuum particles fall to earth at same rate through the electromagnetic interaction (or photon exchange) process, no matter how the virtual masseons combine to give the familiar virtual particles. This process works like a <u>microscopic principle of equivalence</u> for falling virtual particles, with the same action occurring for virtual particles as for large falling masses.

The properties of the masseon particle is elaborated in section 7 (the masseon may be the unification particle sought out by physicist, in which case it will have other properties to do with the other forces of nature). For now, note that the masseon also carries the fundamental unit of electric charge as well. This fundamental unit of electric charge turns out to be the source of inertia for all matter according to Quantum Inertia. By postulating the existence of the masseon particle (which is the fundamental unit of 'mass charge' as well as 'electrical charge') all the quantum vacuum virtual particles accelerate at the same rate with respect to an observer on the surface of the earth. We have postulated the existence of a fundamental "low level gravitational mass charge" of a particle, which results from the graviton particle exchange process similar to the process found for electrical charges. This 'mass charge' is not affected when a particle achieves relativistic velocities, so we can state that 'low level mass charge' is an absolute constant. For particles accelerated to relativistic speeds, a high relative velocity between the source of the force and the receiving mass affects the ordinary measurable inertial mass, as we have seen (in accordance to Einstein's mass-velocity formula).

### (A-7) Summary of the Basic Mass Definitions in EMQG

EMQG proposes three different mass definitions for an object:

(1) INERTIAL MASS is the measurable mass defined in Newton's force law F=MA. This is considered as the absolute mass in EMQG, because it results from force produced by the relative (statistical average) acceleration of the charged virtual particles of the quantum vacuum with respect to the charged particles that make up the inertial mass. The virtual particles of the quantum vacuum become Newton's absolute reference frame. In special relativity this mass is equivalent to the rest mass.

(2) GRAVITATIONAL MASS is the measurable mass involved in the gravitational force as defined in Newton's law $F=GM_1M_2/R^2$. This is what is measured on a weighing scale. This is also considered as absolute mass, and is almost exactly the same as inertial mass.

(3) LOW LEVEL GRAVITATIONAL 'MASS CHARGE', which is the origin of the pure gravitational force, is defined as the force that results through the exchange of graviton particles between two (or more) quantum particles. This type of mass analogous to 'electrical charge', where photon particles are exchanged between electrically charged particles. Note: This force is very hard to measure because it is masked by the background quantum vacuum electromagnetic force interactions, which dominates over the graviton force processes.

These three forms of mass are <u>not</u> necessarily equal! We have seen that the inertial mass is almost exactly the same as gravitational mass, but not perfectly equal. All quantum mass particles (fermions) have all three mass types defined above. Note that bosons (only photons and gravitons are considered here) have only the first two mass types. This means that photons and gravitons transfer momentum, and <u>do</u> react to the presence of inertial frames and to gravitational fields, but they do not emit or absorb gravitons. Gravitational fields effect photons, and this is linked to the concept of space-time curvature, described in detail later (section 9). It is important to realize that gravitational fields deflect photons (and gravitons), but not by force particle exchanges directly. Instead, it is due to a scattering process (described later).

To summarize, both the photon and the graviton do not carry low level 'mass charge', even though they both carry inertial and gravitational mass. The graviton exchange particle, although responsible for a major part of the gravitational mass process, does not itself carry the property of 'mass charge'. Contrast this with conventional physics, where the photon and the graviton both carry a non-zero mass given by $M=E/C^2$. According to this reasoning, the photon and the graviton both carry mass (since they carry energy), and therefore both must have 'mass charge' and exchange gravitons. In other words, the graviton particle not only participates in the exchange process, it also undergoes further exchanges while it is being exchanged! This is the source of great difficulty for canonical quantum gravity theories, and causes all sorts of mathematical renormalization problems in the corresponding quantum field theory. Furthermore, in gravitational force interactions with photons, the strength of the force (which depends on the number of gravitons exchanged with photon) varies with the energy that the photon carries! In modern physics, we do not distinguish between inertial, gravitational, or low level 'mass charge'. They are assumed to be equivalent, and given a generic name 'mass'. In EMQG, the photon and graviton carry measurable inertial and gravitational mass, but neither particle carries the 'low level mass charge', and therefore do not participate in graviton exchanges.

We must emphasize that gravitons do not interact with each other through force exchanges in EMQG, just as photons do not interact with each other with force exchanges in QED. Imagine if gravitons did interact with other gravitons. One might ask how it is possible for a graviton particle (that always moves at the speed of light) to emit graviton particles that are also moving at the speed of light. For one thing, this violates the principles of special relativity theory. Imagine two gravitons moving in the same direction at the speed of light, which are separated by a distance d, with the leading graviton called 'A' and the lagging graviton called 'B'. How can graviton 'B' emit another graviton (also moving at the speed of light) that becomes absorbed by graviton 'A' moving at the speed of light? As we have seen, these difficulties are resolved by realizing that there are actually three different types of mass. There is measurable inertial mass and measurable gravitational mass, and low level 'mass charge' that cannot be directly measured. Inertial and gravitational masses have already been discussed and arise from different physical circumstances, but in most cases give identical results. However, the 'low level mass charge' of a particle is defined simply as the force existing between two identical particles due to the exchange of graviton particles only, which are the vector bosons of the gravitational force. Low level mass charge is not directly measurable, because of the complications due to the electromagnetic forces that are present simultaneously from the virtual particles.

It would be interesting to speculate what the universe might be like if there were no quantum vacuum virtual particles present. Bearing in mind that the graviton exchange process is almost identical to the photon exchange process, and bearing in mind the complete absence of the electromagnetic component in gravitational interactions, the universe would be a very strange place indeed. We would find that large masses would fall faster than smaller masses, just as a large positive electric charge would 'fall' faster than a small positive charge towards a very large negative charge. There would be no inertia as we know it, and basically no force would be required to accelerate or stop a large mass.

**(A-8) The Quantum Field Theory of the Masseon and Graviton Particles**

EMQG addresses gravitational force, inertia, and electromagnetic forces only, and the weak and strong nuclear forces are excluded from consideration. EMQG is based on the idea that all elementary matter particles must get their quantum mass numbers from combinations of just one fundamental matter (and corresponding anti-matter particle), which has just one fixed unit or quanta of mass that we call the 'masseon' particle. This fundamental particle generates a fixed flux of gravitons that are exchanged during gravitational interactions. The exchange process is not affected by the state of motion of the masseon (as you might expect from the special relativistic variation of mass with velocity). We also purpose that nature does <u>not</u> have two completely different long-range forces, for example gravity and electromagnetism. Instead we believe that there exists an almost perfect symmetry between the two forces, which is hidden from view because of the mixing of these two forces in all measurable gravitational interactions. In EMQG the graviton and photon exchange process are found to be essentially the same, except for the strength of the force coupling (and a minor difference in the treatment of positive and negative masses discussed later). EMQG treats graviton exchanges by the same successful methods developed for the behavior of photons in QED. The dimensionless coupling constant that governs the graviton exchange process is what we call '$\beta$' in close analogy with the dimensionless coupling constant '$\alpha$' in QED, where $\beta \approx 10^{-40} \alpha$.

As we stated, EMQG requires the existence of a new fundamental matter particle called the 'masseon' (and a corresponding 'anti-masseon' particle), which are held together by a new unidentified strong force. Furthermore, EMQG requires that masseons and anti-masseons emit gravitons analogous with the electrons and anti-electrons (positrons) which emit photons in QED. Virtual masseons and anti-masseons are created in equal amounts in the quantum vacuum as virtual particle pairs. A masseon generates a fixed flux of graviton particles with wave functions that induce attraction when absorbed by another masseon or anti-masseon; and an anti-masseon generates a fixed flux of graviton particles with an opposite wave function (anti-gravitons) that induces repulsion when absorbed by another masseon or anti-masseon. A graviton is its own anti-particle, just as a photon is its own antiparticle. This process is similar to, but not identical to the photon exchange processes in QED for electrons of opposite charge. In QED, an electron produces a fixed flux of photon particles with wave functions that induces repulsion when absorbed by another electron, and induces attraction when absorbed by a positron. A positron produces a fixed flux of photon particles with wave functions that induces attraction when absorbed by another electron, and induces repulsion when absorbed by a positron. From this it can be seen that if two sufficiently large pieces of anti-matter can be fabricated which are both electrically neutral, they will be found to repel each other gravitationally! Thus anti-matter can actually be thought of as 'negative' mass (-M), and therefore negative energy. This grossly violates the equivalence principle.

These subtle differences in the exchange process in QED and EMQG produce some interesting effects for gravitation that are not found in electromagnetism. For example, a large gravitational mass like the earth does not produce vacuum polarization of virtual particles from the point of view of 'mass-charge' (unlike electromagnetism). In gravitational fields, all the virtual masseon and anti-masseon particles of the vacuum have a net average statistical acceleration directed downwards towards a large mass. This produces a net downward accelerated flux of vacuum particles (acceleration vectors only) that effects other masses immersed in this flux.

In contrast to this, an electrically charged object <u>does</u> produce vacuum polarization. For example, a negatively charged object will cause the positive and negative (electrically charged) virtual particles to accelerate towards and away, respectively from the negatively charged object. Therefore, there is no energy contribution to other real electrically charged test particles placed near the charged object from the vacuum particles, because the electrically charged vacuum particles contribute equal amounts of force from both the upward and downward directions. The individual electrical forces from the vacuum cancel out to zero.

In gravitational fields, the vacuum particles are responsible for the principle of equivalence, precisely because of the lack of vacuum polarization due to gravitational fields. Recall that 'masseon' particles of EMQG are equivalent to the 'parton' particle concept that was introduced by the authors of reference 5 concerning HRP Quantum Inertia. Recall that the masseons and anti-masseons also carry one quantum of electric charge of which there are two types; positive and negative charges. For example masseons come in positive and negative electric charge, and anti-masseons also come in positive and negative charge. A single charged masseon particle accelerating at 1g sees a certain fixed amount of inertial force generated by the virtual particles of the quantum vacuum. In a gravitational field of 1g, a single charged masseon particle on the surface of the earth sees the same quantum vacuum electromagnetic force. In other words, from the vantage point of a masseon particle that makes up the total mass, the virtual particles of the quantum vacuum look exactly the same from the point of view of motion and forces whether it is in an inertial reference frame or in a gravitational field. We propose a new universal constant "i" called the 'inertion', which is defined as the inertial force produced by the action of virtual particles on a single (real charged) masseon particle undergoing a relative acceleration of 1 g. This force is the lowest possible quanta of inertial mass. All other masses are fixed integer combinations of this number. This same constant 'i' is also the lowest possible quanta of gravitational force.

The electric charge that is carried by the electron, positron, quark and anti-quark originate from combinations of masseons, which is the fundamental source of the electrical charge. This explains why a fixed charge relationship exists between the quarks and the leptons, which belong to different families in the standard model. For example, according to the standard model, 1 proton charge precisely equals 1 electron charge (opposite polarity), where the proton is made of 3 quarks. This precise equality arises from the fact that charged masseon particles are present in the internal structure of both the quarks and the electrons (and every other mass particle).

The mathematical renormalization process is applied to particles to avoid infinities encountered in Quantum Field Theory (QFT) calculations. This is justified by postulating a high frequency cutoff of the vacuum processes in the summation of the Feynman diagrams. Recall that QED is formulated on the assumption that a perfect space-time continuum exists. In EMQG, a high frequency cutoff is essential because space is quantized as 'cells', specified by Cellular Automata (CA) theory. In CA theory there is quantization of space as cells. If particles are sufficiently close enough, they completely lose their identity as particles in CA theory, and QFT does not apply at this scale. Since graviton exchanges are almost identical to photon exchanges, we suspect that EMQG is also renormalizable as is QED, with a high frequency cutoff as well. This has not been proven yet. The reason that some current quantum gravity theories are not renormalizable boils down to the fact that the graviton is assumed to be the only boson involved in gravitational interactions. The graviton must therefore exhibit all the characteristics of the gravitational field, including space-time curvature.

In EMQG, the photon exchange and graviton exchange process is virtually identical in its basic nature, which shows the great symmetry between these two forces. As a byproduct of this, the quantum vacuum becomes 'neutral' in terms of gravitational 'mass charge', as the quantum vacuum is known to be neutral with respect to electric charge. This is due to an equal number of positive and negative electrical charged virtual particles and 'gravitational charged' virtual particles created in the quantum vacuum at any given time. This in turn is due to the symmetrical masseon and anti-masseon pair creation process. (EMQG does not resolve the problem of why the universe was created with an apparent imbalance of real ordinary matter and anti-matter mass particles.)

This distortion of the acceleration vectors of the quantum vacuum 'stream' serves as an effective 'electromagnetic guide' for the motion of nearby test masses (themselves consisting of masseons) through space and time. This 'electromagnetic guide' concept replaces the 4D space-time geodesics (which is the path that light takes through curved 4D space-time) that guide light and matter in motion. Because the quantum vacuum virtual particle density is quite high, but not infinite (at least about $10^{90}$ particles/m$^3$), the quantum vacuum acts as a very effective and energetic guide for the motion of light and matter.

## (A-9) Introduction to 4D Space-Time Curvature and EMQG

The physicist A. Wheeler once said that: "space-time geometry 'tells' mass-energy how to move, and mass-energy 'tells' space-time geometry how to curve". In EMQG, this statement must be somewhat revised on the quantum particle level to read: large mass-energy concentrations (consisting of quantum particles) exchanges gravitons with the immediate surrounding virtual particles of the quantum vacuum, causing a downward acceleration (of the net statistical average acceleration vectors) of the quantum vacuum particles. This downward acceleration of the virtual particles of the quantum vacuum 'tells' a nearby test mass (also consisting of real quantum particles) how to move electromagnetically, by the exchange of photons between the electrically charged, and falling virtual particles of the quantum vacuum and the electrical charged, real particles inside the test mass. This new view of gravity is totally based on the ideas of quantum field theory, and thus acknowledging the true particle nature of both matter and forces. It is also shows how nature is non-geometric when examined on the smallest of distance scales, where Riemann geometry is now replaced solely by the interactions of quantum particles existing on a kind of quantized 3D space and separate time on the CA.

Since this downward accelerated stream of charged virtual particles also affects light or real photons and the motion of real matter (for example, matter making up a clock), the concept of space-time must be revised. For example, a light beam moving parallel to the surface of the earth is affected by the downward acceleration of charged virtual particles (electromagnetically), and moves in a curved path. Since light is at the foundation of the measurement process as Einstein showed in special relativity, the concept of space-time must also be affected near the earth by this accelerated 'stream' of virtual particles. Nothing escapes this 'flow', and one can imagine that not even a clock is expected to keep the same time as it would in far space. As a result, a radically new picture of Einstein's curved space-time concept arises from these considerations in EMQG.

The variation of the value of the net statistical average (directional) acceleration vector of the quantum vacuum particles from point to point in space (with respect to the center of a massive object) guides the motion of nearby test masses and the motion of light through electromagnetic means. This process leads to the 4D space-time metric curvature concept of general relativity. With this new viewpoint, it is now easy to understand how one can switch between accelerated and gravitational reference frames. Gravity can be made to cancel out inside a free falling frame (technically at a point) above the earth because we are simply taking on the same net acceleration as the virtual particles at that point. In this scenario, the falling reference frame creates the same quantum vacuum particle background environment as found in an non-accelerated frame, far from all gravitational fields. As a result, light travels in perfectly straight lines when viewed by a falling observer, as specified by special relativity.

Thus in the falling reference frame, a mass 'feels' no force or curvature as it would in empty space, and light travels in straight lines (defined as 'flat' space-time). Thus the mystery as to why different reference frames produce different space-time curvature is solved in EMQG. It is interesting that in an accelerated rocket 4D space-time curvature is also present, but now is caused by another mechanism; the accelerated motion of the floor of the rocket itself. In other words, the space-time curvature, manifesting itself as the path of curved light, is really caused by the accelerated motion of the observer! The observer (now in a state of acceleration with respect to the vacuum), 'sees' the accelerated virtual particle motion in his frame. Furthermore, the motion appears to him to be almost exactly the same as if he were in an equivalent gravitational field. This is why the space-time curvature appears the same in both a gravitational field and an equivalent accelerated frame. These differences between accelerated and gravitational frames imply that equivalence is not a basic element of reality, but merely a result of different physical processes, which happen to give the same results. In fact, equivalence is <u>not</u> perfect!

According to EMQG, all metric theories of gravity, including general relativity, have a limited range of application. These theories are useful only when a sufficient mass is available to significantly distort the virtual particle motion surrounding the mass; and only where the electromagnetic interaction dominates over the graviton processes (or where the graviton flux is not too large). For precise calculation of gravitational force interactions of small masses, EMQG requires that the gravitational interaction be calculated by adding the specific Feynman diagrams for both photon and graviton exchanges. Thus, the use of the general relativistic Schwarzchild Metric for spherical bodies (even if modified by including the uncertainty principle) is totally useless for understanding the gravitational interactions of elementary particles. The whole concept of space-time 'foam' is incorrect according to EMQG, along with all the causality problems associated with this complex mathematical concept.

## (A-10) Space-Time Curvature is a Pure Virtual Particle Quantum Vacuum Process

4D Minkowski curved space-time takes on a radically new meaning in EMQG, and is no longer a basic physical element of our reality. Instead, it is merely the result of quantum particle interactions alone. The curved space-time of general relativity arises strictly out of the interactions between the falling virtual particles of the quantum vacuum near a massive object and a nearby test mass. The effect of the falling quantum vacuum acts somewhat like a special kind of "Fizeau-Fluid" or media, that effects the propagation of light; and also effects clocks, rulers, and measuring instruments. Fizeau demonstrated in the middle 1850's that moving water varies the velocity of light propagating through it.

This effect was analyzed mathematically by Lorentz. He used his newly developed microscopic theory for the propagation of light in matter to study how photons move in a flowing stream of transparent fluid. He reasoned that photons would change velocity by frequent scattering with the molecules of the water, where the photons are absorbed and later remitted after a small time delay. This concept is discussed in detail in section 9.3.

If Einstein himself had known about the existence of the quantum vacuum when he was developing general relativity theory, he may have deduced that space-time curvature was caused by the "accelerated quantum vacuum fluid". He was aware of the work by Fizeau, but was unaware of the existence of the quantum vacuum. After all, Einstein certainly realized that clocks were not expected to keep time correctly when immersed in an accelerated stream of water! We show mathematically in this paper that the quantity of space-time curvature near a spherical object predicted by the Schwarzchild metric is identical to the value given by the 'Fizeau-like' scattering process in EMQG.

In EMQG when we find an accelerated vacuum disturbance, there follows a corresponding space-time distortion (including the possibility of gravitational waves for dynamic accelerated disturbances). We have seen that both accelerated and gravitational frames qualify for the status of curved 4D space-time (although caused by <u>different</u> physical circumstances). We have found that in EMQG there exists two, separate but related space-time coordinate systems. First, there is the familiar global four dimensional relativistic space-time of Minkowski, as defined by our measuring instruments, and is designated by the x,y,z,t in Cartesian coordinates. The amount of 4D space-time curvature is influenced by accelerated frames and by gravitational frames, which is the cause of the accelerated state of the quantum vacuum.

Secondly there is a kind of a quantized absolute space, and separate time as required by cellular automaton theory. Absolute space consists of an array of numbers or cells C(x,y,z) that changes state after every new clock operation $\Delta t$. C(x,y,z) acts like the absolute three dimensional pre-relativistic space, with a separate absolute time that acts to evolve the numerical state of the cellular automata. The CA space (and separate time) is not affected by any physical interactions or directly accessible through any measuring instruments, and currently remains a postulate of EMQG. Note that EMQG absolute space does not correspond to Newton's idea of absolute space. Newton postulated the existence of absolute space in his work on inertia. He realized that absolute space was required in order to resolve the puzzle of what reference frame nature uses to gauge accelerated motion. In EMQG, this reference frame is <u>not</u> the absolute quantized cell space (which is unobservable), but instead consists of the net average state of acceleration of the virtual particles of the quantum vacuum with respect to matter (particles). A very important consequence of the existence of absolute quantized space and quantized time (required by cellular automaton theory) is the fact that our universe must have a maximum speed limit!

## (A-11) THE BASIC POSTULATES OF EMQG

Here is a summary of the basic postulates of EMQG. Reference 1 gives a much more complete description of the postulates and their consequences. Note that we do not include Einstein's principle of equivalence as one of our basic postulates. This is because equivalence is not a fundamental principle. Instead equivalence is simply a consequence of quantum particle interactions. The basic postulates of EMQG are:

### POSTULATE #1:    CELLULAR AUTOMATA

The universe is a vast cellular automaton computation, which has an inherently quantized absolute 3D space consisting of 'cells', and absolute time. The numeric information in a cell changes state through the action of the numeric content of the immediate neighboring cells (26 neighbors) and the local mathematical rules, which are repeated for each and every cell. The action of absolute time (through clock cycles) synchronizes the state transition of all the cells. The number of 'clock cycles' elapsed between the change of the numeric state on the CA is a measure of the absolute time elapsed. The cells are interconnected (mathematically) to form a simple 3D geometric CA. Matter, forces, and motion are the end result of information changing in the cells as absolute time progresses. Gravity, motion, and any other physical process do **not** effect low-level absolute 3D space and absolute time in any way. Photons propagate in the simplest possible manner on the CA. Photons simply shift from cell to adjacent cell on each and every 'clock cycle' in a given direction. This rate represents the maximum speed that information can be moved during a CA 'clock cycle'. The quantization scale is not known yet, but it must be much finer than the Plank Scale for distance and time.

### POSTULATE #2:    GRAVITON-MASSEON PARTICLES

The masseon is the most elementary form of matter (or anti-matter), and carries the lowest possible quanta of low level, gravitational 'mass charge'. The masseon carries the lowest possible quanta of positive gravitational 'mass charge', where the low level gravitational 'mass charge' is defined as the (probability) fixed rate of emission of graviton particles in close analogy to electric charge in QED. Gravitational 'mass charge' is a fixed constant and analogous to the fixed electrical charge concept. Gravitational 'mass charge' is **not** governed by the ordinary physical laws of *observable* mass, which appear as 'm' in the various physical theories, including Einstein's special relativity mass-velocity relationship: $E=mc^2$ or $m = m_0 (1 - v^2/c^2)^{-1/2}$. Masseons simultaneously carry a positive gravitational

'mass charge', and either a positive or negative electrical charge (defined exactly as in QED). Therefore we conclude that masseons also exchange photons with other masseon particles. Masseons are fermions with half integer spin, which behave according to the rules of quantum field theory. Gravitons (which are closely analogous to photons) have a spin of one (not spin two, as is commonly thought), and travel at the speed of light. Anti-masseons carry the lowest quanta of negative gravitational 'mass charge'. Anti-masseons also carry either positive or negative electrical charge, with electrical charge being defined according to QED. An anti-masseon is always created with an ordinary masseon in a particle pair as required by quantum field theory (specifically, the Dirac equation). The anti-masseon is the negative energy solution of the Dirac equation for a fermion, where now the **mass** is taken to be 'negative' as well, in clear violation of the principle of equivalence. Another important property exhibited by the graviton particle is the **principle of superposition**. This property works the same way as for photons. The action of the gravitons originating from all sources acts to yield a net vector sum for the receiving particle. EMQG treats graviton exchanges by the same successful methods developed for the behavior of photons in QED. The dimensionless coupling constant that governs the graviton exchange process is what we call '$\beta$' in close analogy with the dimensionless coupling constant '$\alpha$' in QED, where $\beta \approx 10^{-40} \alpha$.

## POSTULATE #3: QUANTUM THEORY OF INERTIA

The property which Newton called the inertial mass of an object, is caused by the vacuum resistance to acceleration of all the individual, electrically charged masseon particles that make up the mass. This resistance force is caused by the electromagnetic force interaction (where the details of this process are unknown at this time) occurring between the electrically charged virtual masseon/anti-masseon particle pairs created in the surrounding quantum vacuum, and all the real masseons particles making up the accelerated mass. Therefore inertia originates in the photon exchanges with the electrically charged virtual masseon particles of the quantum vacuum. The total inertial force $F_i$ of a mass is simply the sum of all the little forces $f_p$ contributed by each of the individual masseons, where the sum is: $F_i = (\Sigma f_p) = MA$ (Newton's law of inertia).

## POSTULATE #4: PHOTON FIZEAU-LIKE SCATTERING IN THE VACUUM

Photons have an absolute, fixed velocity resulting from its special motion on the CA, where photons simply shift from cell to adjacent cell on every CA 'clock cycle'. This 'low level' photon velocity (measured in CA absolute space and time units) is *much higher* (by an unknown amount) than the observed light velocity of 300,000 km/sec. This is because a photon traveling in the vacuum (in an inertial frame) takes on a path through the quantum vacuum, that is the end result of a vast number of electromagnetic scattering processes with the surrounding electrically charged virtual particles. Each scattering process introduces a small random delay in the subsequent remission of the photon, and results in a cumulative reduction in the velocity of photon propagation. Real photons that travel near a large mass like the earth, take a path through the quantum vacuum that is the end result of a large number of electromagnetic scattering processes with the falling (statistical average) electrically charged virtual particles of the quantum vacuum. The resulting path is one where the photons maintain a net statistical average acceleration of zero with respect to the electrically charged virtual particles of the quantum vacuum, through a process that is very similar to the Fizeau scattering of light through moving water. Through very frequent absorption and re-emission (which introduces a small delay) by the accelerated charged virtual particles of the quantum vacuum, the apparent light velocity assumes an accelerated value with respect to the center of mass *in absolute* CA space and time units. (Note: ***The light velocity is still an absolute constant when moving in between virtual particles, and is always created at this fixed constant velocity dictated by the CA rules***). The accelerated virtual particles of the quantum vacuum that appears in gravitational and accelerated reference frames can be viewed as a special Fizeau-like vacuum fluid. This fluid effects the motion of matter and light in the direction of the fluid acceleration, which is ultimately responsible for 4D space-time curvature.

### (A-12) Experimental Verification of EMQG Theory

EMQG proposes several new experimental tests that give results that differ from the conventional general relativistic physics, and can thus be used to verify the theory.

(1) EMQG opens up a new field of physics, which we call anti-matter gravitational physics. We propose that if two sufficiently large pieces of anti-matter are manufactured to allow measurement of the mutual gravitational interaction (with a torsion balance apparatus for example), then the gravitational force will be found to be repulsive! The force will be equal in magnitude to $-GM^2/r^2$ where M is the mass of each of the anti-matter masses, r is their mutual separation, and G is Newton's gravitational constant. This is a ***gross violation*** of the principle of equivalence, since in this case, $M_i = -M_g$, instead of being strictly equal. Antimatter that is accelerated in far space has the same inertial mass '$M_i$' as ordinary matter, but when interacting gravitationally with another antimatter mass it is repelled ($M_g$). Note that the earth will attract bulk anti-matter because of the large abundance of gravitons originating from the earth of the type that induce attraction. This means that no violation of equivalence is expected for anti-matter dropped on the earth, where anti-matter falls normally. However, an antimatter earth will repel a nearby antimatter mass. Recent attempts at measuring earth's gravitational force on anti-matter (for example on anti-protons) will not reveal any deviation from equivalence, according to EMQG. However, if there were two large identical masses of matter and anti-matter close to each other, there would be no gravitational force existing between them because of the balance of "positive and negative" masses, for example equal numbers of gravitons that induce attraction and repulsion. This gravitational system is considered gravitationally 'neutral' as is the quantum vacuum, which is also gravitationally neutral.

(2) For an extremely large test mass and a very small test mass that is dropped simultaneously on the earth (in a vacuum), there will be an *extremely small* difference in the arrival time of the masses on the surface of the earth in slight violation of the principle of equivalence. This effect is on the order of $\approx \Delta N \times \delta$, where $\Delta N$ is the difference in the number of masseon particles in the two masses, and $\delta$ is the ratio of the gravitational to electric forces for one masseon. This experiment is very difficult to perform on the earth, because δ is extremely small ($\approx 10^{-40}$), and ΔN cannot be made sufficiently large. To achieve a difference of $\Delta N = 10^{30}$ masseons particles between the small and large mass requires dropping a molecular-sized cluster and a large military tank simultaneously in the vacuum in order to give a measurable deviation. Note that for ordinary objects that might seem to have a large enough difference in mass (like dropping a feather and a tank), the difference in arrival time would be obscured by background interference, and possibly by quantum effects like the Heisenberg uncertainty principle which restrict the accuracy of arrival time measurements.

(3) If gravitons can be detected by the invention of a graviton detector/counter in the far future, then there will be experimental proof for the violation of the strong principle of equivalence. The strong equivalence principle states that all the laws of physics are the same for an observer situated on the surface of the earth as it is for an accelerated observer at 1 g. The graviton detector will find a tremendous difference in the graviton count in these two cases. This is because gravitons are vastly more numerous here on the earth. Thus a detector can manufactured with an indicator that distinguishes between whether an observer is in an inertial frame or in a gravitational frame. This of course is a *gross violation* of the strong equivalence principle.

(4) Since the gravitational mass of an object has a strong electrical force component, mass measurements near the earth might be disrupted experimentally by manipulating some of the electrically charged virtual particles of the nearby quantum vacuum through electromagnetic means. If a rapidly fluctuating magnetic field (or rotating magnetic field) is produced directly under a mass it might effect the instantaneous charged virtual particle spectrum and disrupt the tiny electrical forces for many of the masseons in the mass. This may reduce the measured gravitational (and inertial masses) of a test mass. In a sense this device would act like a primitive and weak "anti-gravity" machine.

The virtual particles are constantly being "turned-over" in the vacuum at different rates, with the high frequency virtual particles (and therefore, the high-energy virtual particles) being replaced the quickest. If a magnetic field is made to fluctuate fast enough so that it does not allow the new electrically charged virtual particle pairs to replace the old and smooth out the disruption, the spectrum of the virtual particles in the vicinity may be altered. According to conventional physics, the energy density of virtual particles is infinite, which means that all frequencies of virtual particles are present. In EMQG there is an upper cut-off to the frequency, and therefore the highest energy according to the Plank's law: E=hυ, where υ is the frequency that a virtual particle can have. We can state that the smallest wavelength that a virtual particle can have is about $10^{-35}$ meters, e.g. the plank wavelength (or a corresponding maximum Plank frequency of about $10^{43}$ hertz for very high velocity (≈c) virtual particles). Unfortunately for our "anti-gravity" device, it is technologically impossible to disrupt these highest frequencies. Recall that according to the uncertainty principle, the relationship between energy and time is: $\Delta E \, \Delta t > h/(2\pi)$. This means that the high frequency end of the spectrum consists of virtual particles that "turns-over" the fastest. To give maximum disruption to a significant percentage of the high frequency virtual particles require magnetic fluctuations on the order of at least $10^{20}$ cycles per seconds. Therefore only lower frequencies virtual particles of the vacuum can be practically affected in the future, and only small changes in the measured mass can be expected with today's technology.

As a result of this we conclude that the higher the frequency the greater the mass loss. Recent work on the Quantum Hall Effect by Laughlin on fractional electron charge suggests that, under the influence of a strong magnetic field, electrons might move in concert with swirling vortices created in the 2D electron gas. This leads to the possibility that this 'whirlpool' phenomenon also holds for the virtual particles of the quantum vacuum under the influence of a strongly fluctuating magnetic field. These high-speed whirlpools might disrupt the high frequency end of the distribution of electrically charged virtual particles into small pockets. Therefore, there might be a greater mass loss under these circumstances, an idea this is very speculative at this time. Experiments that report mass reduction associated with rapidly rotating superconducting magnets, which generate high frequency rotating magnetic fields are inconclusive at this time. Reference 6 gives an excellent and detailed review of the various experiments on reducing the gravitational force with superconducting magnets.

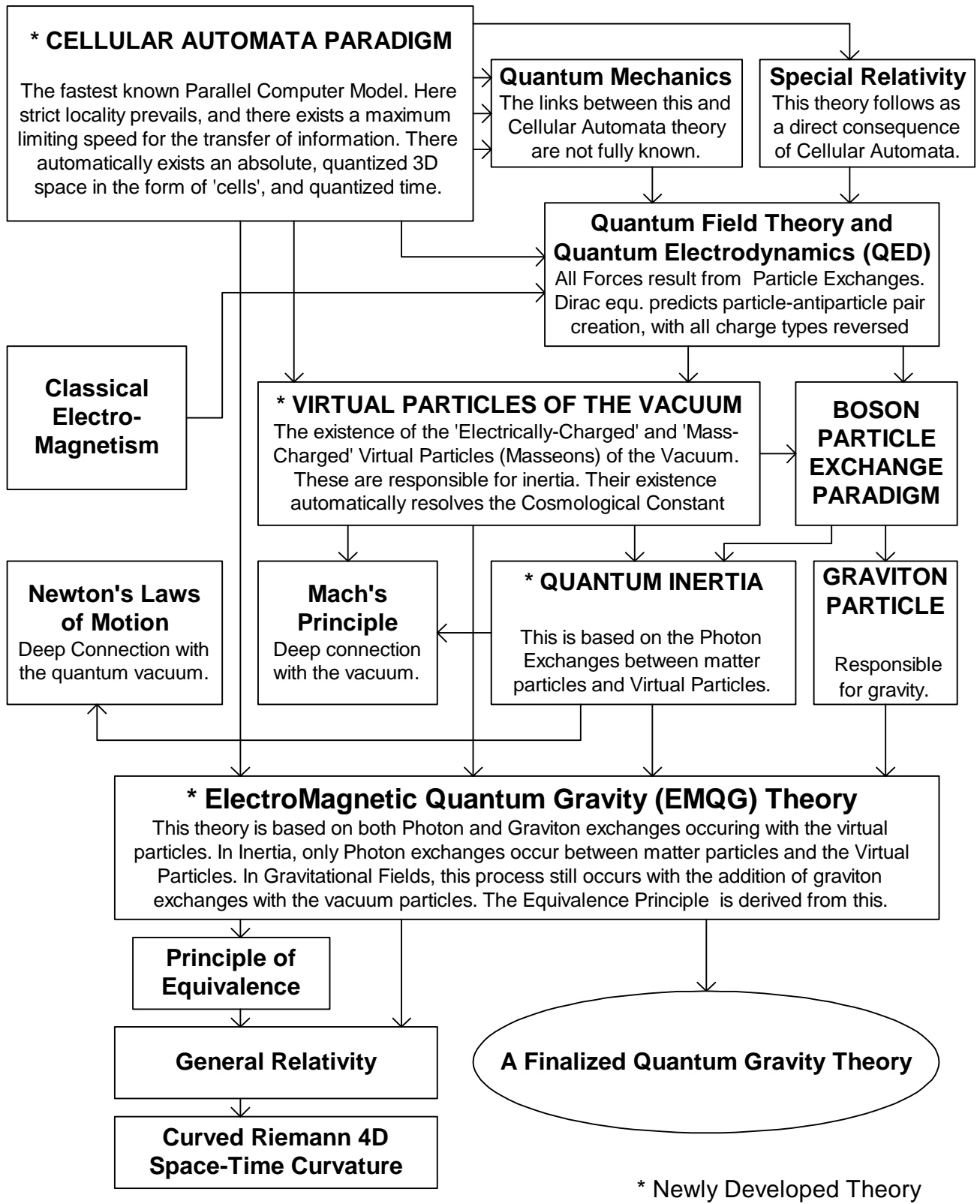

**Figure #1 - BLOCK DIAGRAM OF RELATIONSHIP OF CA AND EMQG WITH PHYSICS**

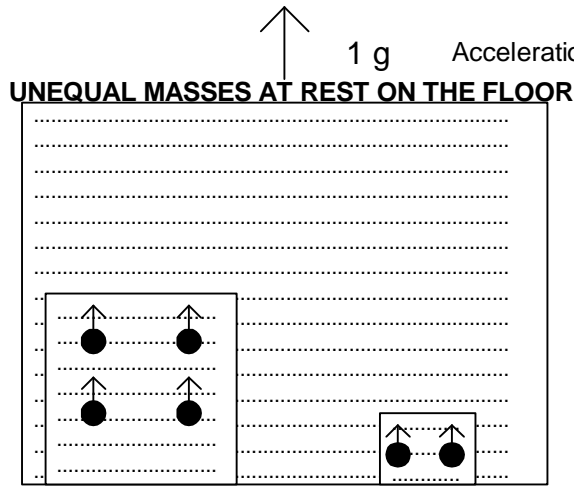

**Figure 2A** - Masses '2M' and 'M' at rest on the floor of the rocket

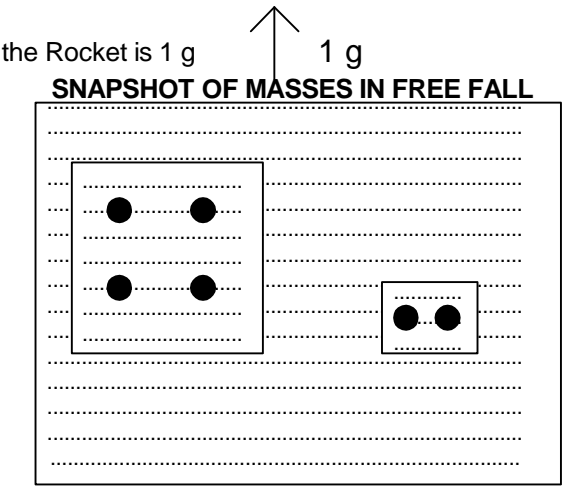

**Figure 2B** - Masses '2M' and 'M' in free fall inside of a rocket

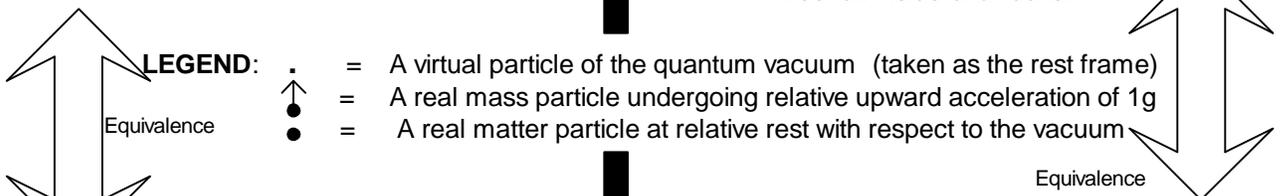

**LEGEND**:
- . = A virtual particle of the quantum vacuum (taken as the rest frame)
- ↑● = A real mass particle undergoing relative upward acceleration of 1g
- ● = A real matter particle at relative rest with respect to the vacuum

Equivalence

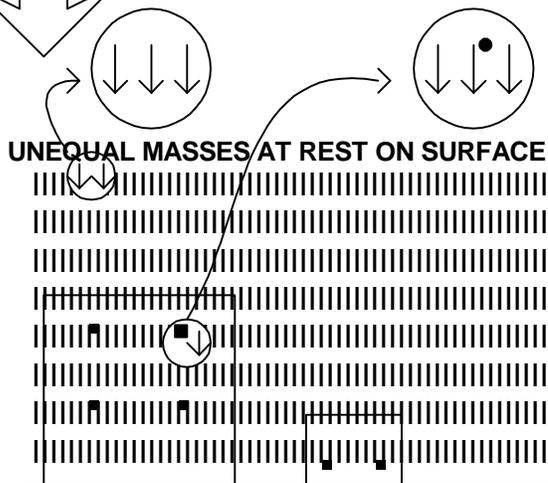

**Figure 2C** - Masses '2M' and 'M' at rest on Earth's surface

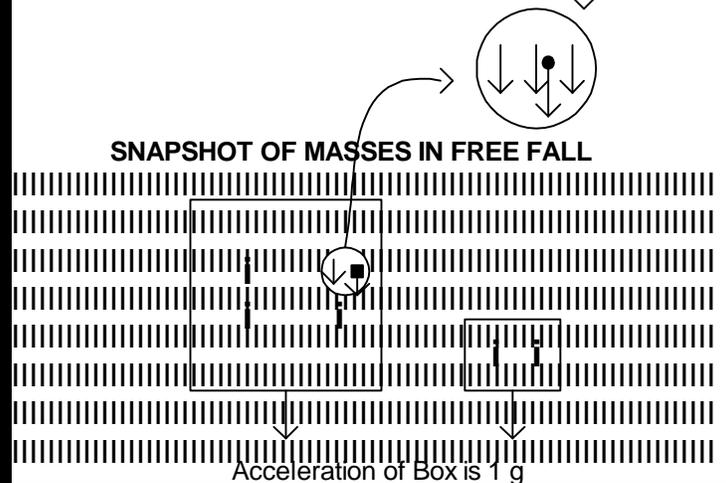

Acceleration of Box is 1 g

Surface of the Earth where gravity produces a 1 g acceleration

**Figure 2D** - Masses '2M' and 'M' in free fall above the Earth

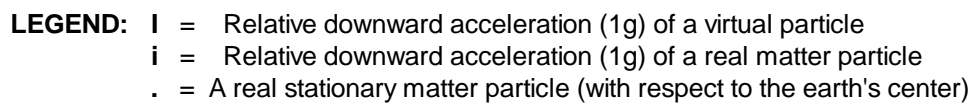

**LEGEND:**
- l = Relative downward acceleration (1g) of a virtual particle
- i = Relative downward acceleration (1g) of a real matter particle
- . = A real stationary matter particle (with respect to the earth's center)

**Figure #2 - SCHEMATIC DIAGRAM OF THE PRINCIPLE OF EQUIVALENCE**